\documentclass[%
 reprint,
 amsmath,amssymb,
 aps,
 prl,
]{revtex4-1}

\usepackage[utf8]{inputenc}
\usepackage{graphicx}
\usepackage{hyperref}
\hypersetup{colorlinks = true, linkcolor = [rgb]{0.19411,0.51882,0.667058}, urlcolor = [rgb]{0.125490,0.29542,0.1647058}, citecolor = [rgb]{0.75882,0.37411,0.14117}}
\usepackage{amsmath}
\usepackage{blkarray}
\usepackage{amsfonts}
\usepackage{amssymb}
\usepackage{siunitx}
\usepackage{braket}
\usepackage{import}
\usepackage{dsfont}
\usepackage{color,soul}

\begin{document}

\title{Large baseline optical imaging assisted by single photons and linear quantum optics}

\author{Marta Maria Marchese}
\author{Pieter Kok}
\affiliation{Department of Physics \& Astronomy, the University of Sheffield, Hounsfield Road, Sheffield, S3 7RH, UK.}

\begin{abstract}
In this work, we show that by combining quantum metrology and networking tools, it is possible to extend the baseline of an interferometric optical telescope and thus improve diffraction-limited imaging of point source positions. The quantum interferometer is based on single-photon sources, linear optical circuits, and efficient photon number counters. Surprisingly, with thermal (stellar) sources of low photon number per mode and high transmission losses across the baseline, the detected photon probability distribution still retains a large amount of Fisher information about the source position, allowing for a significant improvement in the resolution of positioning point sources, on the order of 10\;$\mu$as. Our proposal can be implemented with current technology. In particular, our proposal does not require experimental optical quantum memories.
\end{abstract}

\maketitle

\noindent
\color{black}
Telescopes can be improved in several ways:
improving the signal-to-noise ratio \cite{pearce2017optimal,howard2019}, employing (quantum) designs to achieve super-resolution \cite{kolobov2007quantum,tsang2016quantum,paur2016achieving,tsang2019resolving,lupo2020,zanforlin2022,Raymer22}, or extending the baseline to create larger apertures \cite{Dravins16,akiyama2019first,boyer2022,Czupryniak2021,Raymer2022}.
The resolving power of an optical imaging system can be defined in terms of the minimum resolvable angle $\theta_{\rm min}$, which depends on the source's wavelength $\lambda$ and on the telescope's aperture size. 
Larger aperture sizes give better resolving power. In astronomy, large baseline interferometric telescopes have become an established technique to create highly resolved images. 
One of the most notable results achieved with this technique is the first-ever radio image of a black hole~\cite{akiyama2019first}, produced by a combination of many telescopes that formed a single imaging system with an aperture the size of the Earth. The receivers measured the phase and amplitude of the radio signals, and combined them into an interferometric image.
\color{black}

For optical signals, the frequency is too high to measure the phase and amplitude of the incoming light directly, and the incoming signals must be made to interfere physically. This places a limit on how far receivers can be placed apart, since the transmission of the signal in optical fibres or light tunnels is lossy \cite{townsend1993single}. A solution involving teleporting photons from the receiver to the interferometric setup demonstrates how quantum information technology can overcome this limitation using optical quantum memories \cite{gottesman2012longer,khabiboulline2019optical}. Recently, a large-baseline quantum telescope was proposed that makes use of photonic quantum memories and error correction to protect the weak optical signal from transmission losses \cite{PhysRevLett.129.210502}.

\begin{figure}[t!]
    \centering
    \includegraphics[width=1\columnwidth]{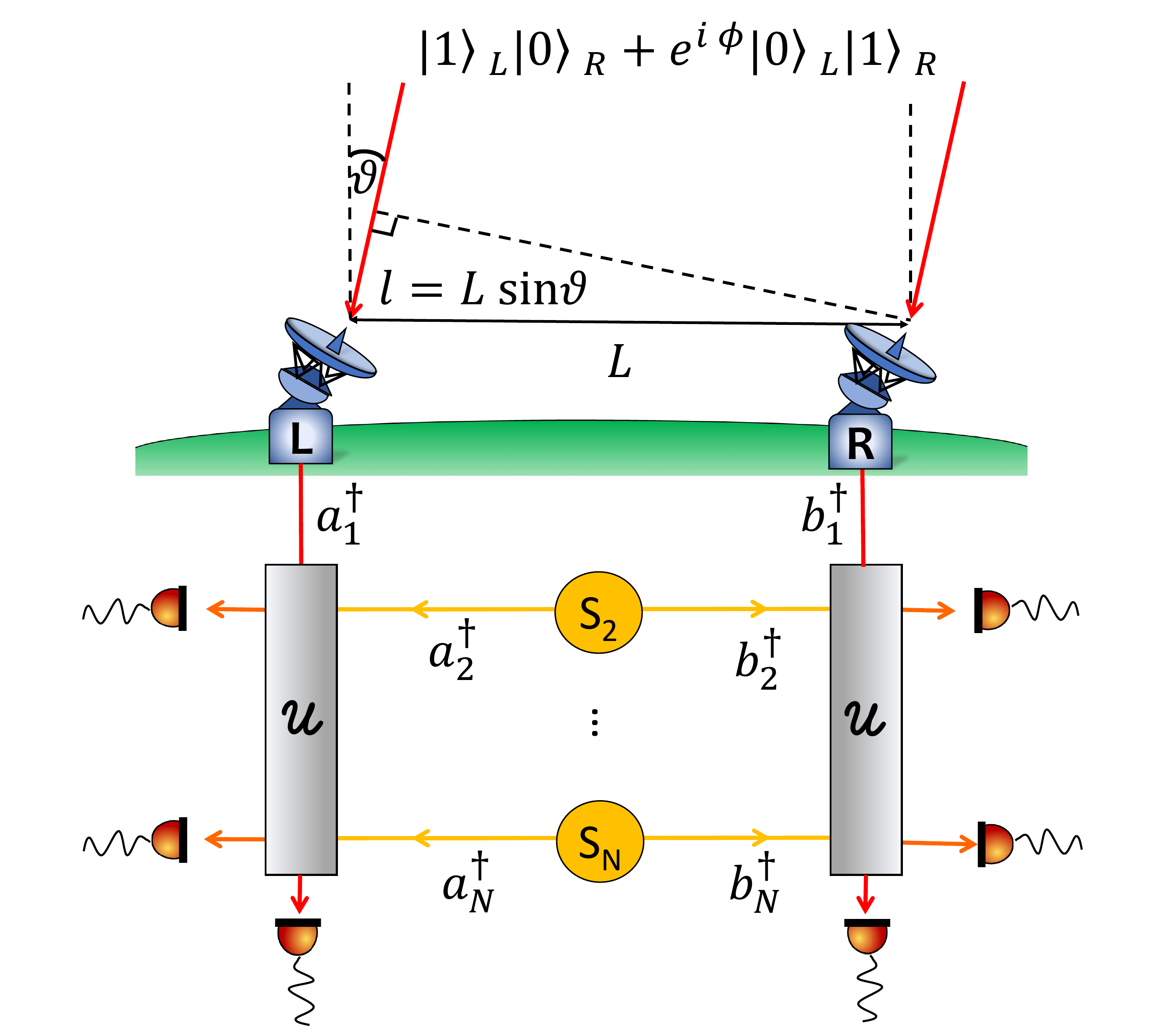}
    \caption{Interferometric telescope consisting of two receivers separated by a distance $L$. The light from the faraway star arrives at both sites in modes $a_1$ and $b_1$ with a relative phase shift $\phi=k l$, determined by the path length difference $l$ and wave number $k$. A quantum network of single-photon sources $S_j$ ($j=2\ldots N$), linear optical circuits $\mathcal{U}$, and photon counting is used to perform interferometric measurements at each site.}
    \label{fig:system}
\end{figure}

While future quantum technologies can deliver major improvements to large baseline optical telescopes, near-future telescopes must rely on more readily available technologies. Here, we propose a large-baseline optical telescope that employs single photon sources~\cite{kaneda2019high,kennard2013chip,fulconis2005high,silverstone2014chip}, low-loss optical fibres, linear optical interferometry~\cite{wang2020integrated,carolan2015universal}, and high-efficiency photon number counters \cite{mcmillan2013optical,paesani2020near}. Our proposal uses multiple single-photon sources and optical quantum Fourier transform (QFT) circuits, shown in Fig.~\ref{fig:system}, which is a linear optical circuit where a photon entering any input mode is equally likely to to appear in any output mode. Here, we demonstrate that establishing coherence across the baseline using multiple single photon sources and beam splitters improves the resolution of the telescope, even when the transmission losses along the baseline are high.

Our setup is as follows: a simple interferometric telescope employs two receivers, A and B, a distance $L$ apart. A distant star-source emits incoherent light that can be described as a single mode with large transverse coherence reaching the telescope at an angle $\theta$. At any given optical frequency, the distant star emits a weak incoherent signal such that we can assume at most single photon events with probability $\epsilon$. The state of a single photon entering the two receivers is given by 
\begin{equation}
    |\psi\rangle_1=\dfrac{|1\rangle_A|0\rangle_B+e^{i\phi}|0\rangle_A|1\rangle_B}{\sqrt{2}},
    \label{eq:staridealstate}
\end{equation}
where $|0\rangle_{L,R}$ and $|1\rangle_{L,R}$ indicates the zero and one photon state at each receiver, and $\phi$ is the relative phase shift between the two receivers. It arises from the additional distance $l=L\sin\theta$ in the light-path at one of the receivers, and $\phi=k l$, where $k$ is the wave number of the source. For distant star-sources a very good approximation is $\phi = k L \theta$. {\color{black}The quantum Fisher information for $\phi$ in the state $|\psi\rangle_1$ is 1 (see Supplementary Material).}

In an ideal scenario there is no loss, and the interferometric measurement can be simply performed by recombining the photon through a beam-splitter. However, a major challenge is the physical transportation of the photon from one site to the other. In a realistic dissipative scenario the photon loss thus limits the baseline $L$. In our proposal, the star photon ($S_1$) does not travel a large distance from the receiver, and incurs minimal loss. Instead, single photons generated in ground-based sources $S_j$ ($j=2\ldots N$) are sent to the two receivers, $A$ and $B$, in the single-photon Bell state
\begin{equation}
    |\psi_j\rangle=\dfrac{|1\rangle_A|0\rangle_B+|0\rangle_A|1\rangle_B}{\sqrt{2}}, 
    \label{eq:labstate}
\end{equation}
which can be produced using a 50:50 beam splitter. At each receiver, we let these photons interfere with the optical mode from the starlight in a QFT circuit, and the output modes are measured in highly efficient photon counting detectors \cite{hadfield2009single,esmaeil2021superconducting,natarajan2012superconducting,hadfield2006quantum,miller2003tungsten}. The information of the signal we intend to measure is contained in the correlations between the detectors at the receivers. By using multiple ground-based photon sources, we aim to overcome the large transmission losses between the receivers.

We treat the imaging protocol above as a quantum parameter estimation problem for $\phi$ \cite{paris2009quantum,giovannetti2006quantum,giovannetti2011advances,demkowicz2009quantum,knott2016local}, which is directly related to $\theta$. The ultimate precision with which it is possible to measure $\phi$ is given by the Cram\'er-Rao bound~\cite{helstrom1976quantum}, which represents a lower bound to the variance $(\delta\phi)^2$ of an estimator of $\phi$, given the knowledge of the quantum mechanical state $\rho(\phi)$. {\color{black} For unbiased estimators, t}he bound is given by the inverse of the Fisher information $F(\phi)$ associated with the state
\begin{equation}
    (\delta\phi)^2\geq\dfrac{1}{M F(\phi)},
    \label{eq:variance}
\end{equation}
where $M$ is the number of independent measurements. The Fisher information
\begin{equation}
    F(\phi)=\int \text{d} x\; p(x|\phi)\left(\dfrac{\partial \ln p(x|\phi)}{\partial \phi}\right)^2,
    \label{eq:Fisher}
\end{equation}\\
is given in terms of the probabilities $p(x|\phi)$ of measuring the outcome $x$ when the parameter has fixed value $\phi$.

First, we consider a single ground-based photon and a single star photon. If both photons are detected by the same receiver, then $\phi$ is at most a global phase in the quantum state, and the measurements will not reveal any information about $\phi$. Only when the two photons are detected at different receivers do we gain information about $\phi$. This limits the Fisher information to $F_2(\phi) = \frac12$. To increase the Fisher information, we can increase the number of ground-based photons to $N-1$, which means that the probability of measuring all photons at the same receiver becomes $1/N$, and the resulting Fisher information is bounded by 
\begin{equation}
    F_N \leq 1-\dfrac{1}{N}\, .
\label{eq:fisherN_NOLoss}
\end{equation}
At this point, we have assumed no losses in the transmission, and we did not make assumptions about the precise interferometer at both receivers.

Next, we introduce the model for describing the quantum telescope with $N-1$ ground-based photons. We will describe the lossless case first.  The total initial state is
\begin{equation}
    |\psi\rangle_{\rm{tot}}^{\rm{in}}=|\psi_1\rangle\otimes|\psi_2\rangle...|\otimes|\psi_N\rangle\, ,
    \label{eq:idealtotalinitialstate}
\end{equation}
where the first photon comes from the star-source  $S_1$ and its initial state $|\psi_1\rangle$ is given by Eq.~\eqref{eq:staridealstate}, and the other $N-1$ photons are in the state $|\psi_j\rangle$, given in Eq.~\eqref{eq:labstate}.
In terms of creation and annihilation operators $\{a^\dagger_n,a_n\}$ and $\{b^\dagger_n,b_n\}$ for each photon mode, the state becomes
\begin{equation}
\begin{split}
    |\psi\rangle_{\rm{tot}}^{\rm{in}}=&
    \left(\dfrac{1}{2}\right)^{\frac{N}{2}}\prod_{n=1}^N\left(a^\dagger_n+e^{i\phi \delta_{n,1}}b^\dagger_n\right)|0\rangle,
    \label{eq:initialstateN}
\end{split}
\end{equation}
where $\delta_{n,1}$ is the Kronecher delta indicating that the relative phase shift enters just in the mode of the star-source $S_1$, and $|0\rangle$ is the vacuum state for all the modes. For the interferometry between the star-photon and the lab-photons we choose a linear transformation $\mathcal{U}$ that implements a QFT~\cite{reck1994experimental}. For each of the $N$ input modes $a_n$ on the right
\begin{equation}
    a_n^{\dagger}=\dfrac{1}{\sqrt{N}}\sum_{k=1}^N\omega^{n k}a_k^\dagger,
    \label{eq:QFT}
\end{equation}
and analogously for the $N$ input modes $b_n$ on the left. The output state $|\psi\rangle_{\rm{tot}}^{\rm{out}}$ is then
\begin{equation}
    |\psi\rangle_{\rm{tot}}^{\rm{out}}= \left(\dfrac{1}{2N}\right)^{\frac{N}{2}}\prod_{n=1}^N\sum_{k=1}^N\omega^{nk}\left(a_k^\dagger+e^{i\phi\delta_{n,1}}b_k^\dagger\right)|0\rangle.
\end{equation}
The probability of measuring a given configuration $\mathbf{d}$ of $N$ photons across the $2N$ detectors is
\begin{equation}
    P_{\mathbf{d}}(\phi)=|\langle \mathbf{d}|\psi\rangle_{\rm{tot}}^{\rm{out}}|^2,
\end{equation}
where $|\mathbf{d}\rangle=|d_1,d_2,...,d_j,...,d_{2N}\rangle$ is the measured state with $d_i\in\{1,N\}$ photons at the detector $i$.
We recast the Fisher information in terms of these probabilities, as
\begin{equation}
     F_N(\phi)=\sum_{ \mathbf{d}}^{\sigma_N} P_{ \mathbf{d}}(\phi)\left(\dfrac{\partial \text{ln}P_{ \mathbf{d}}(\phi)}{\partial \phi}\right)^2,
     \label{eq:FisherN}
\end{equation}
where the sum runs over all the possible configurations $\sigma_N$ of $N$ photons distributed across $2N$ detectors. For the lossless case, $F_N(\phi)$ is shown in the Supplementary Material (see Fig.~\ref{fig:FisherNOloss}).

\begin{figure}
    \centering
    \includegraphics[width=1\columnwidth]{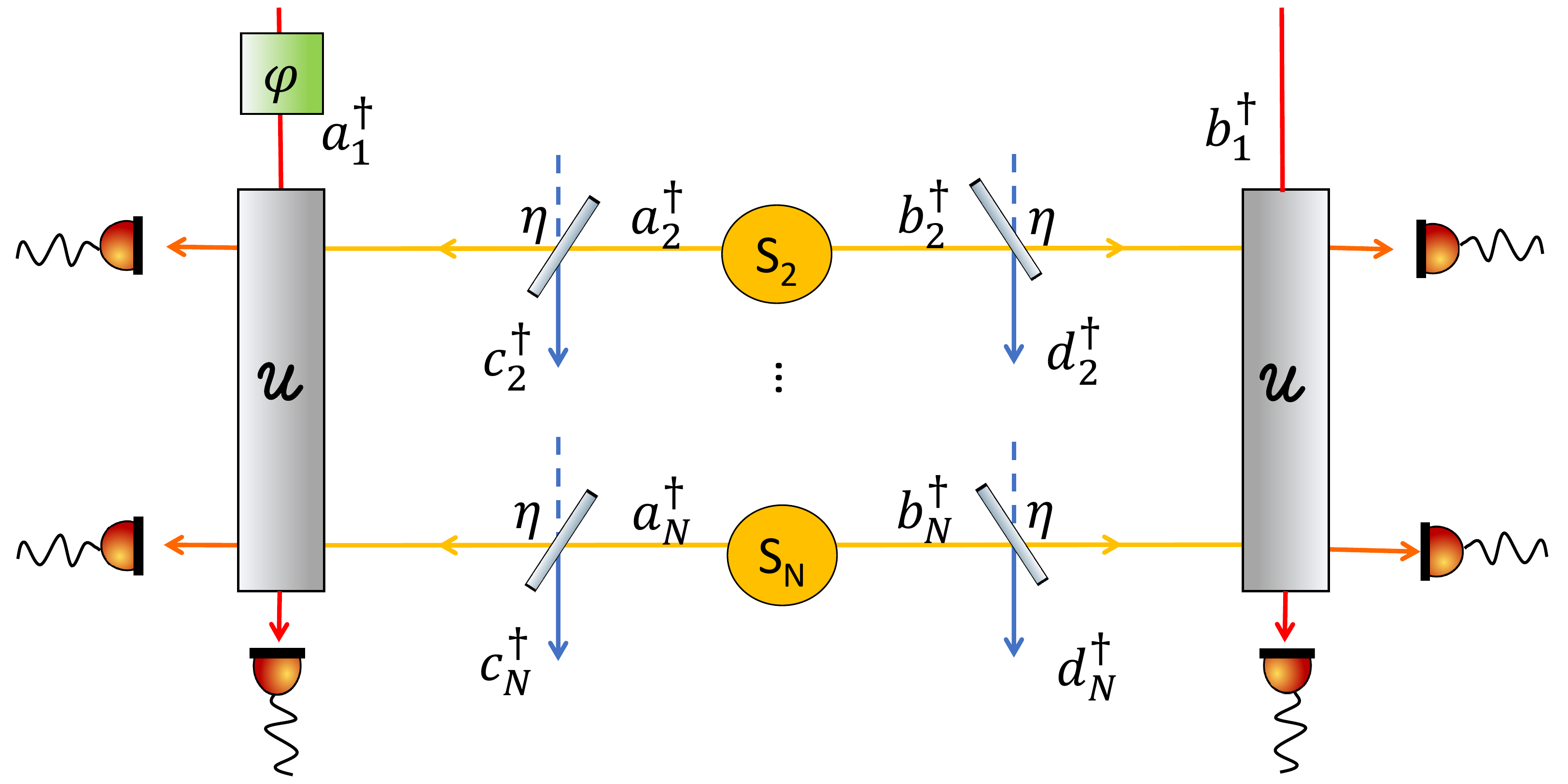}
    \caption{The presence of transmission loss in the ground-based photons can be modelled with beam splitters with transmissivity $\eta$. These will mix the input modes $\{a^\dagger_j,b^\dagger_j\}$, for the ground-based photons, with vacuum input modes, represented by the dashed blue lines, which are subsequently traced over. Each photon mode will be subject to loss with probability $p=1-\eta^2$. The star photon coming from source $S_1$ is represented by mode operators $\{a^{\dagger}_1,b^{\dagger}_1\}$. A phase shift $\varphi$ is included in mode $a_1$ to allow measurements at optimal phase differences.}
    \label{fig:loss}
\end{figure}

The ground-based photons are subject to transmission losses, which can be modelled using a beam splitter in the transmission line. This is shown in Fig.~\ref{fig:loss}. The transmission probability amplitude of a photon in an optical fibre is given by $\eta=e^{-L/4L_0}$, where we use that each photon travels over a length $L/2$, and $L_0$ is the fibre attenuation length---assumed identical for all transmission lines. The transformation for the $j$-photon mode operators is thus
\begin{equation}
\begin{split}
    &a_j^\dagger=\eta a^\dagger_j+\sqrt{1-\eta^2}c_j^\dagger\\
    &b_j^\dagger=\eta b^\dagger_j+\sqrt{1-\eta^2}d^\dagger_j,
\end{split}
\end{equation}
where $\{c^\dagger_j,c_j\}$ and $\{d^\dagger_j, d_j\}$ are the creation and annihilation operators for the vacuum fields respectively on the left and right side.

For now, we consider the presence of exactly one star-photon. The state of the optical modes in Eq.~\eqref{eq:initialstateN} at the QFT circuits in the presence of loss becomes 
\begin{equation}
\begin{split}
    &\left(\dfrac{1}{2}\right)^{N/2}\left(a^\dagger_1+e^{i\phi}b^\dagger_1\right)\otimes\\
    &\prod_{n=2}^N\left[\sqrt{1-p}\left(a^\dagger_n+b^\dagger_n\right)+\sqrt{p}\left(c_n^\dagger+d_n^\dagger\right)\right]|0\rangle,
\end{split}
\end{equation}
where $p=1-\eta^2$ is the probability of losing a single photon. 

The number of detected photons $d$ is no longer equal to $N$; it runs in the interval $[1,N]$. The Fisher information will be modified, since it will include the probabilities for partial photon detection. In the Supplementary Material we show that this can be rewritten as the weighted sum of Fisher information contributions corresponding to partial photon detection
\begin{equation}
     F_N^{\rm{loss}}=\sum_{k=0}^{N-1} (1-p)^{N-1-k}(p)^k\binom{N-1}{k} F'_{N-k}.
    \label{eq:fisherLOSS}
\end{equation}
where $F'_{N-k}$ is the Fisher information for $D=N-k$ photon detected, $k$ is the number of photons lost.  

\begin{figure}
    \centering
    \includegraphics[width=1\columnwidth]{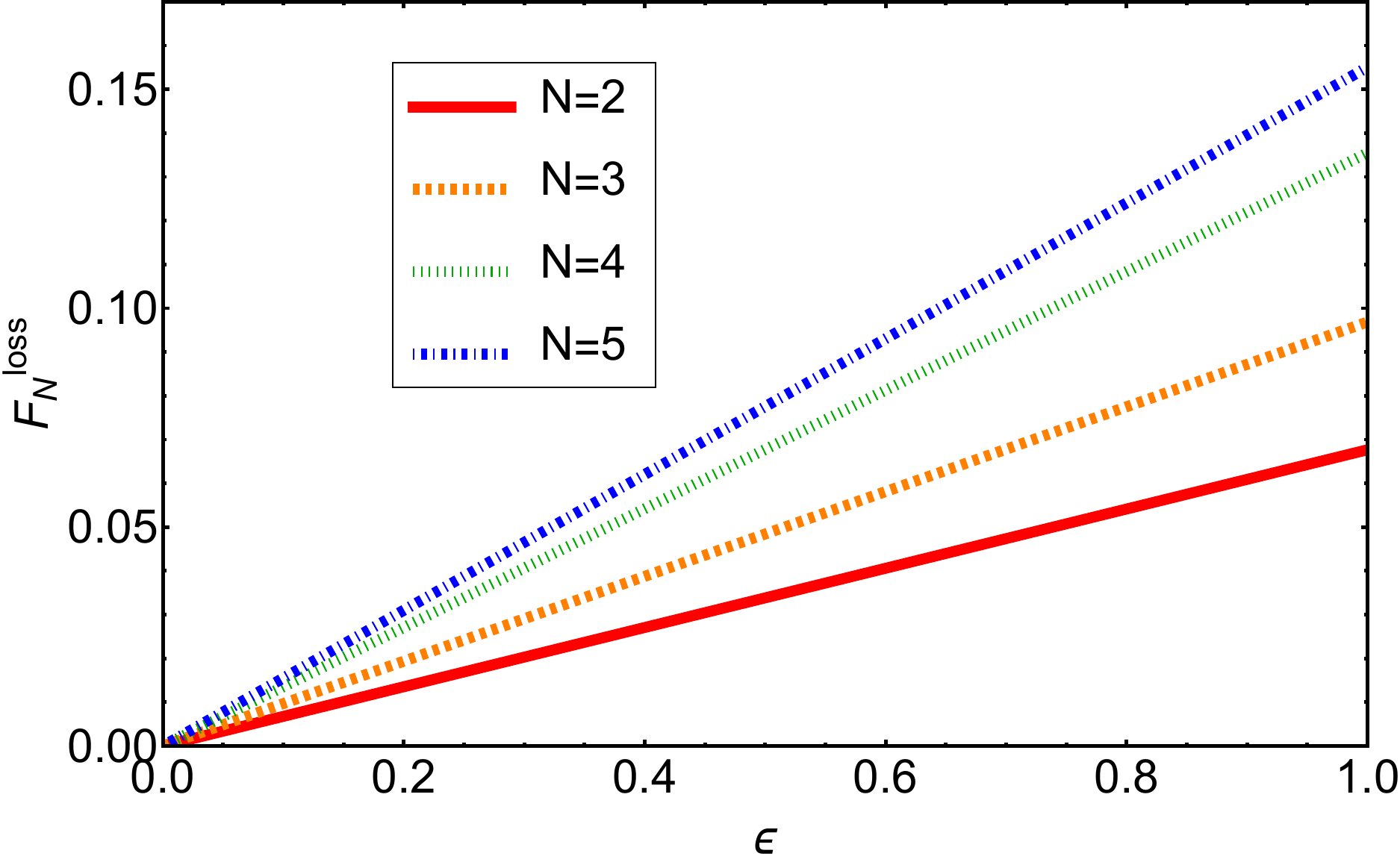}
    \caption{The Fisher information in presence of loss scales linearly with the star emission rate $\epsilon$. The curves have been obtained for the optimal values of relative phase shift $\varphi_{\rm{opt}}$ and baseline length $\alpha_{\rm{opt}}$ reported in Table~\ref{table:results}.}
    \label{fig:linearityfisher}
\end{figure}

Next, we consider the case where the starlight is in a thermal state at an optical frequency with a rate of photon emission $\epsilon$ much smaller than 1 \cite{mandel1995optical}. Therefore, the density operator for the initial state of the star light can be well approximated as
$\rho_S=(1-\epsilon)\rho_0+\epsilon\rho_1$,
where  $\rho_1$ is the one-photon state from Eq.~\eqref{eq:idealtotalinitialstate}, and $\rho_0=|0\rangle\langle 0|$ is the zero-photon state. 
The probability of detecting $d$ photons is given by the sum of two probabilities as
\begin{equation}
    P_T(\mathbf{d})=(1-\epsilon) P_A(\mathbf{d})+\epsilon P_B(\mathbf{d}),
    \label{eq:probtot}
\end{equation}
where $P_A(\mathbf{d})$ and $P_B(\mathbf{d})$ are the probability distributions for having no star photon and having a star photon, respectively. This means that when detecting $d<N$ photons, there will be two contributions to the total probability distribution, arising from the fact that we can not distinguish whether the reduced photon number is due to lossy transmission of the ground-based photons or an absence of the star photon. 

The Fisher information must be modified as follows {\color{black}(see Supplementary Materials)}
\begin{equation}
    F(\phi)=\sum_{\mathbf{d}}^{\sigma_D}\dfrac{1}{P_T(\mathbf{d})}\left(\dfrac{\partial P_T(\mathbf{d})}{\partial \phi}\right)^2,\label{eq:FisherLossD}
\end{equation}
and by using Eq.~\eqref{eq:probtot} we obtain
\begin{equation}
      F(\phi)=\sum_{\mathbf{d}}^{\sigma_D}\dfrac{\epsilon^2}{(1-\epsilon) P_A(\mathbf{d})+\epsilon P_B(\mathbf{d})}\left(\dfrac{\partial P_B(\mathbf{d})}{\partial \phi}\right)^2\, .
\end{equation}
Here, we used that in the absence of a star photon the probability distribution $P_A(\mathbf{d})$ does not depend on $\phi$, and the contribution to the derivative is zero. We calculated the Fisher information for the cases $N=2,3,4,5$, but these expressions are too large to include here.

\begin{figure}[t!]
    \centering
    \includegraphics[width=0.9\columnwidth]{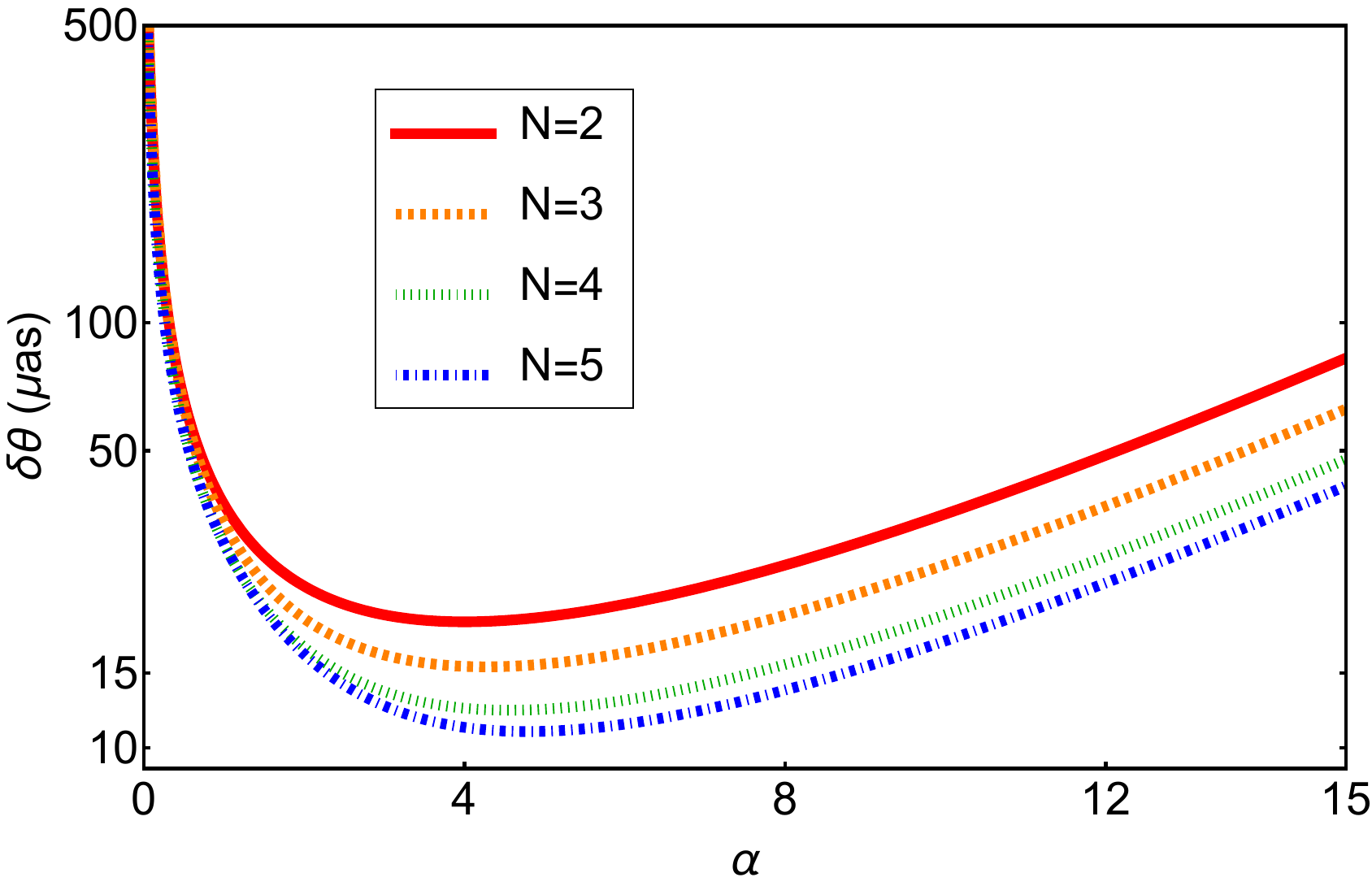}
    \caption{
    The resolution angle $\delta\theta$ in micro-arcseconds, as a function of $\alpha=L/L_0$, the baseline length in units of the attenuation length $L_0$. The curves have been obtained for optical wavelengths $\lambda=628~{\rm{nm}}$, star emission rate $\epsilon=0.01$ and typical attenuation length value $L_0=10~{\rm{km}}$. Different colors correspond to a different total photon number $N$. At short distances, the resolution angle decreases as we increase the baseline; it reaches a minimum value and then increases due to losses becoming predominant over longer distances. We make two observations: (i) the more ground-based photons are employed the lower the minimum of the resolution, and (ii) the minimum is shifted towards larger and larger distance as we increase $N$, allowing for an extension of the baseline.}
    \label{fig:resolution}
\end{figure}

The total Fisher information is modified by the $\epsilon$ factor, which is usually very small. A Fisher information  $F(\phi)$ scaling linearly in $\epsilon$ reflects the reduced rate $\epsilon$ of gaining information about $\phi$. However, when $F(\phi)$ scales with $\epsilon^2$, this would indicate that the transmission losses obscure the absence of a star photon, leading to a much deteriorated metrology protocol and a very large estimation error. Surprisingly, we find that for the cases we examined ($N=2,3,4,5$), the Fisher information scales \emph{linearly} in $\epsilon$, as shown in Fig.~\ref{fig:linearityfisher}. This can be understood by considering that the transmission loss and the absence of a star photon occur in orthogonal optical modes, rendering $P_A(\mathbf{d})$ and $P_B(\mathbf{d})$ quite different. Consequently, our protocol is still able to provide an improvement in the resolution of $\phi$ (and therefore $\theta$).

Finally, we study the resolution of the telescope in terms of the error $\delta\theta$ in the geometrical angle $\theta = \phi/kL$ under the small angle approximation. The mean square error is given by 
\begin{equation}
    (\delta\theta)^2=\dfrac{(\delta\phi)^2}{k^2L^2},
\end{equation}
where $(\delta \phi)^2$ is the variance associated with the relative phase shift, lower bounded by the Fisher information. Therefore, the minimum error on the resolution is given by
\begin{equation}
    (\delta\theta)^2=\left(\dfrac{1}{kL}\right)^2\dfrac{1}{F_N(\phi)}.
    \label{eq:varianceresolution}
\end{equation}
The resolution $\delta\theta$ will be better for higher values of the Fisher information and for large distances between the telescopes, in accordance with the Rayleigh criterion.

The resolution of the telescope will improve when we increase the distance between the receivers, but increased transmission photon loss will deteriorate the resolution. Therefore, we must find the optimal distance between receivers for photon-assisted interferometric imaging. This is shown in Fig.~\ref{fig:resolution}. For light with a wavelength $\lambda=628$~nm and fibre attenuation length $L_0=10$~km, an angular resolution of $19.80$~$\mu$as {\color{black}(micro arcseconds)} is achievable with a single ground-based photon ($N=2$), while four ground-based photons ($N=5$) yields an angular resolution of $10.93$~$\mu$as. For increasing $N$, the best resolution is obtained for larger distances, indicating that the extra ground-based photons help increase the baseline, and hence the resolution. We summarise the optimal distances and the achievable resolution in Table~\ref{table:results}. Moreover, the corresponding optimal relative phase shift $\phi_{\rm{opt}}$ is determined by the orientation of the baseline, and can be set by including an adjustable phase shift at one of the receivers.

\begin{table}[t!]
\centering
\begin{tabular}{p{2cm} p{2cm} p{2cm} p{2cm}} 
 \hline  \hline
 \centering
 $N$ & $\delta\theta_{\rm{min}} $($\mu\rm{as}$) & $\alpha_{\rm{opt}}$ & $\varphi_{\rm{opt}}$ (rad) \\[0.5ex]
 \hline
 \centering
 2 & 19.80 & 3.99 & 0.318\,$\pi$ \\ 
 \centering
 3 & 15.52 & 4.17 & 0.724\,$\pi$ \\
 \centering
 4 & 12.26 & 4.56 & 0.513\,$\pi$ \\
 \centering
 5 & 10.93 & 4.79 & 0.247\,$\pi$\\ [1ex] 
 \hline
\end{tabular}
\caption{The table lists the parameters obtained for different number of photons $N$. The second column shows the minimum resolution $\delta\theta_{\rm{min}}$, while the third and fourth columns report the optimal values of the baseline $\alpha=L_{opt}/L_0$ and the relative phase shift $\varphi_{\rm{opt}}$ for which it is obtained.}
\label{table:results}
\end{table}

{\textit{Conclusions}}- {\color{black}
We addressed the challenge of increasing the resolution of positioning incoherent point sources in interferometric telescopes at optical frequencies by extending the baseline of the telescope. 
We showed that a dramatic imaging resolution improvement can be obtained with a setup that relies solely on current technology (multiple single-photon sources, optical fibres, linear optical circuits and photon number counting detectors), without the need for quantum memories or full-scale quantum repeater networks. This is an example of a useful near- to medium-term quantum technology application beyond quantum key distribution. 

Transmitting multiple single photons across the baseline with high transmission losses extends the numerical aperture to tens of kilometres for optical frequencies, leading to a resolution on the order of 10\;$\mu$as.
One may expect that losing photons in an interferometric measurement will drastically reduce the Fisher information, since we cannot distinguish between photons that are lost in transmission or that were not emitted by the source in the first place. 
However, the Fisher information scales linearly rather than quadratic in $\epsilon\ll1$, which accounts for this unexpected good performance. 
Our results are important for our theoretical understanding of optical interferometers in metrology applications, since they point towards a pathway for loss-tolerant optical quantum metrology where signal photons and auxiliary optical states interfere. It may help extend the range of optical communication without full quantum repeaters.

Finally, our analysis is not limited to ground-based telescopes and apply equally to satellite-based receivers where the single photon sources are distributed through free space. The beam divergence is the main loss mechanism, leading to much larger baselines on the order of 1\,000\;km or a resolution of 50\;nas. 
}

\emph{Acknowledgements}.~
We thank Zixin Huang, Cosmo Lupo, and Francesco Albarelli for valuable discussions and suggestions. This work is funded by the EPSRC Large Baseline Quantum-Enhanced Imaging Networks Grant No. EP/V021303/1, and the EPSRC Quantum Communications Hub, Grant No. EP/M013472/1.

\newpage
\phantom{.}
\newpage

\widetext 

\appendix
{\color{black}
\section{Quantum Fisher information for a single phase}
We show here that the Quantum Fisher information of $\phi$ for the state $|\psi\rangle=\dfrac{|0\rangle+e^{i\phi}|1\rangle}{\sqrt{2}}$ is equal to one.
The Quantum Fisher information
\begin{equation}
    F_Q=4 (\Delta G)^2,
\end{equation}
is expressed in terms of the variance $(\Delta G)^2=\langle G^2\rangle-\langle G\rangle^2$ associated with the generator $G=\sigma_z/2$. Therefore, we can directly compute the variance on the state $|\psi\rangle$ as
\begin{equation}
    (\Delta G)^2=\dfrac{1}{4}\langle \psi|\psi\rangle-\dfrac{1}{4} \left[\left(\dfrac{\langle 0| +e^{-i\phi}\langle 1|}{\sqrt{2}}\right)\left(\dfrac{|0\rangle-e^{i\phi}|1\rangle}{\sqrt{2}}\right)\right]^2=\dfrac{1}{4}.
\end{equation}
It follows that the Quantum Fisher information is $F_Q=1$.
}

\section{From the $2$-photon to $N$-photon case}\label{app:N2}
\begin{figure}
    \centering
    \includegraphics[width=0.5\columnwidth]{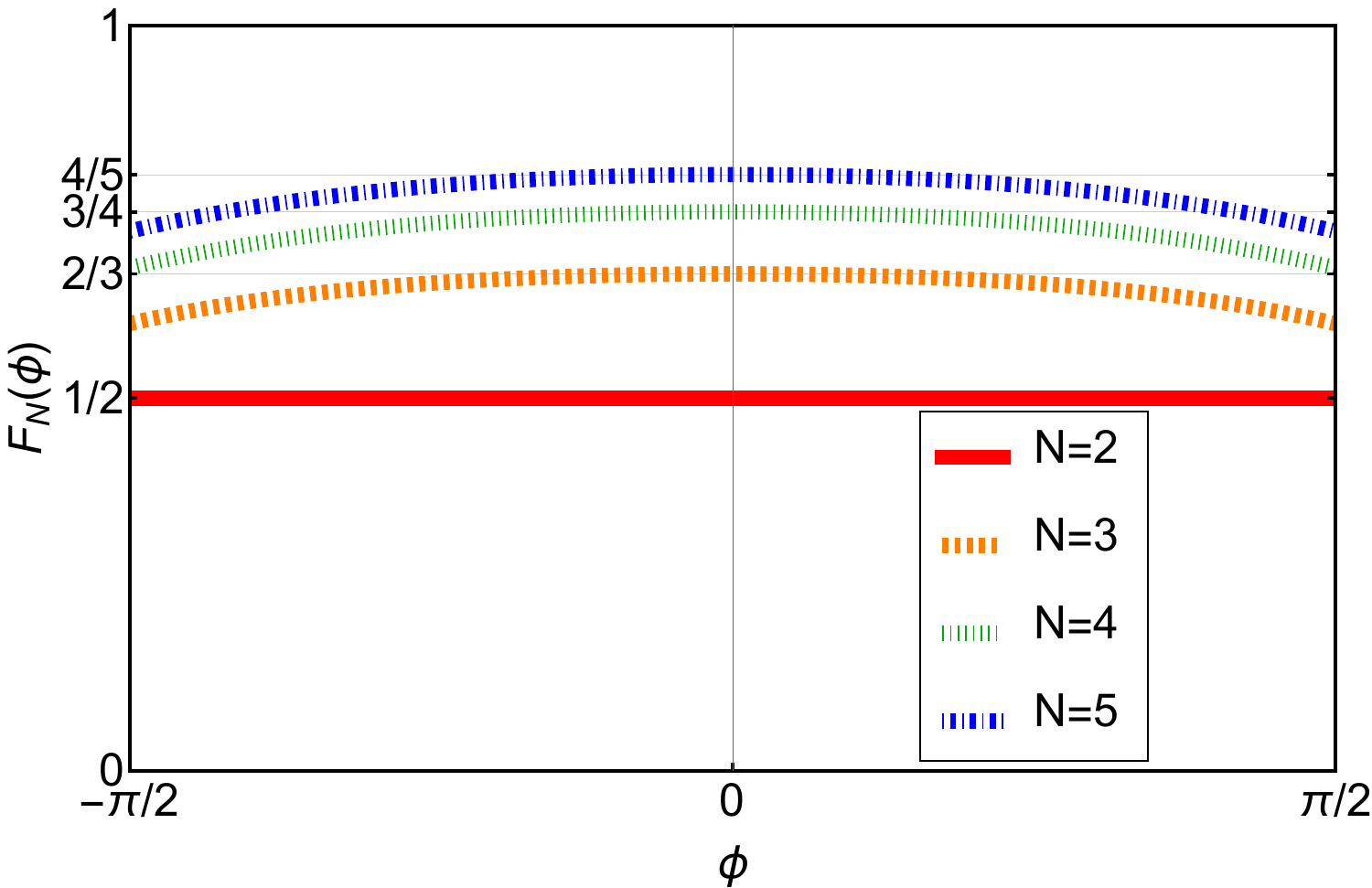}
    \caption{The Fisher information in function of the relative phase shift $\phi$ for different number of photons $N$, in the ideal case where the is no loss. The red curve is the simplest case with 2 photons, where the interferometric measurements are performed with 1 Lab photon; The blue, green and orange functions are obtained with $N=3,4,5$ photons respectively, and they provide a Fisher information that increases with the number of photons as $F_N(\phi)=1-1/N$.}
    \label{fig:FisherNOloss}
\end{figure}

We discuss in detail the simplest scenario consisting of only two photon sources, one  $S_1$ for the star photon, and the other $S_2$ for the ground-based photon. The total initial state is the tensor product of the star photon state, given by Eq.~\eqref{eq:staridealstate}, and the ground-based photon state, from Eq.~\eqref{eq:labstate},
\begin{equation}
    |\psi\rangle_{\rm{tot}}^{\rm{in}}=\left(\dfrac{|10\rangle+e^{i\phi}|01\rangle}{\sqrt{2}}\right)_1\otimes\left(\dfrac{|10\rangle+|01\rangle}{\sqrt{2}}\right)_2,
\end{equation}
where we simplified the notation by omitting the subscripts $\{L,R\}$. \\
It is convenient to rewrite the initial state in term of the creation/annihilation operators, as
\begin{equation}
     |\psi\rangle_{\rm{tot}}^{\rm{in}}= \dfrac{1}{\sqrt{2}}\left(a^\dagger_1+e^{i\phi}b^\dagger_1\right)\otimes \dfrac{1}{\sqrt{2}}\left(a^\dagger_2+b^\dagger_2\right)|0\rangle,
     \label{eq:initialstate2modes}
\end{equation}
where we called $\{a^\dagger_i,a_i\}$ ($\{b^\dagger_i,b_i\}$) the operators for the right (left) states, and the index $i\in\{1,2\}$ indicates either the star source $S_1$ or the ground-based photon source $S_2$; $|0\rangle$ is the total vacuum state.\\
At each site, the star photon modes are mixed with the ground-based photon modes according to the linear QFT evolution from Eq.~\eqref{eq:QFT}. For the case $N=2$ this transformation takes the form of a simple 50:50 beam-splitter, that gives the following outputs for the $a_i$ modes on the right
\begin{equation}
    a_1^{\dagger_{\rm{out}}}=\left(\dfrac{-a_1^{\dagger_{\rm{in}}}+a_2^{\dagger_{\rm{in}}}}{\sqrt{2}}\right)\qquad\text{and}\qquad
     a_2^{\dagger_{\rm{out}}}=\left(\dfrac{a_1^{\dagger_{\rm{in}}}+a_2^{\dagger_{\rm{in}}}}{\sqrt{2}}\right),
\end{equation}
and analogously for the $b_i$ modes on the left.
Therefore, the initial state in Eq.~\eqref{eq:initialstate2modes} will be transformed into the final state
\begin{equation}
     |\psi\rangle_{\rm{tot}}^{\rm{out}}=\left(\dfrac{-a^\dagger_1+a^\dagger_2-e^{i\phi}b^\dagger_1+e^{i\phi}b^\dagger_2}{2}\right)\left(  \dfrac{a^\dagger_1+a^\dagger_2+b^\dagger_1+b^\dagger_2}{2}\right)|0\rangle.
\end{equation}
To compute the Fisher information we need the probabilities
\begin{equation}
P_{\mathbf{d}}(\phi)=|\langle\mathbf{d}|\psi\rangle_{\rm{tot}}^{\rm{out}}|^2,
\end{equation}
for all the configurations $|\mathbf{d}\rangle=|d_1,d_2,d_3,d_4\rangle$, where $d_i\in\{1,2\}$ is the number of detected photons.
By looking at the probabilities, for this simple case, there are ten possible ways of distributing 2 photons among the four detectors. However, only when the 2 photons are detected on different sides we can extract information on $\phi$. This happens for four configurations, that will give probabilities 
\begin{equation}
\small
\begin{split}
   & P_{|1,0,0,1\rangle}(\phi)=P_{|0,1,1,0\rangle}(\phi)=1/8(1-\cos\phi),\\
    &P_{|0,1,0,1\rangle}(\phi)=P_{|1,0,1,0\rangle}(\phi)=1/8(1+\cos \phi).\\
\end{split}\label{eq:2modeProb}
    \end{equation}
All the other configurations will give probabilities that do not depend on $\phi$. In fact, when both photons are detected by the same detector or by different detectors on the same side, the relative phase shift will be just a global phase.
The majority of cases will not bring any contribution to the Fisher information. 
The Fisher information from Eq.~\eqref{eq:Fisher}, in the 2-photon case,  results to be constant
\begin{equation}
    F_2(\phi)=\dfrac{1}{2}.
\end{equation}
The generalization to $N$-photons scenario, where the set-up is endowed with $N-1$ ground-based photon sources, increases the number of probabilities of detecting photons on both sides. This will result into more contributions to the Fisher information and smaller variance.
The plots in Fig.~\ref{fig:FisherNOloss} show the Fisher information in function of the relative phase shift $\phi$ for different number $N$ of photons. As expected, in the ideal no-loss case scenario, the larger number of photons results in a higher Fisher information. For small angles, the Fisher information is approximately constant and it scales with $N$ as $F_N=1-\dfrac{1}{N}$. In the ideal scenario, with no loss, increasing $N$ results in higher values of the Fisher information and lower variances associated to the phase shift angle $\phi$.

\section{Additivity of Fisher information} \label{app:additivityFisher}
We derive an expression for the Fisher information for $N$ photons when the ground-based photons are subjected to transmission losses. This results in a sum of contributions corresponding to the Fisher information of partial photon detection. \\
Consider the case of 1 star photon and $N-1$ ground-based photons in presence of loss. We assume for now that the star photon is not subjected to any loss.  We call $k\in[0,N-1]$ the number of lost photons and $D=N-k$ the number of detected photons. Let's call $\sigma_D$ the set of all the configurations we obtain from measuring $D$ photons. Thus, there will be $N$ cases corresponding to losses of different number of photons, i.e., $\sigma_D\in\{\sigma_1,...,\sigma_N\}$. The two extreme cases are  $\sigma_N$, the set of configurations corresponding to the no-loss scenario, or equivalently all the $N$ photons are detected; and $\sigma_1$ the set of configurations for the case in which all the $N-1$ ground-based photons are lost and only the star photon is detected. 
For a given number $D$ of detected photon, $\sigma_D$ will contain $m_D$ configurations; each configuration $j\in[0,D]$ has a probability $p_{D,j}$
\begin{equation}
        \sigma_D=\{p_{D,1},...,p_{D,j},...,p_{D,m_D}\}.
\end{equation}
The Fisher information from Eq.~\eqref{eq:FisherN} has to be the sum of all the probabilities, thus
\begin{equation}
    F_N^{\rm{loss}}(\phi)=\sum_{D=1}^{N} \sum_{j=1}^{m_D} p_{D,j} \left(\dfrac{\partial \ln{p_{D,j}}}{\partial \phi}\right)^2,
    \label{eq:fisherLOSS1}
\end{equation}
where the sum over $j$ runs over all the possible configurations $m_D$ for $D\in[1,N]$ detected photons, and the other sum accounts for all the possible values of $D$.
The set of probabilities $\sigma_D$ for a given number of detected photons $D$ is not normalized, meaning that the sum does not add up to one, but to a certain value that corresponds to the total probability of detecting $D$ photons 
\begin{equation}
    q_D=\sum_{j=1}^{m_D}p_{D,j}.
\end{equation}
Therefore we rewrite these probabilities in terms of a new set of normalized probabilities
\begin{equation}
    r_{D,j}=\dfrac{p_{D,j}}{q_D}.
\end{equation}
We can use this expression for the Fisher information of Eq.~\eqref{eq:fisherLOSS1} and, by using the logarithmic properties and the fact that the coefficients $q_D$ do not depend on $\phi$, we obtain

\begin{figure}
 \centering
 \begin{minipage}[c]{\columnwidth}
 \centering
        \includegraphics[width=3.0in]{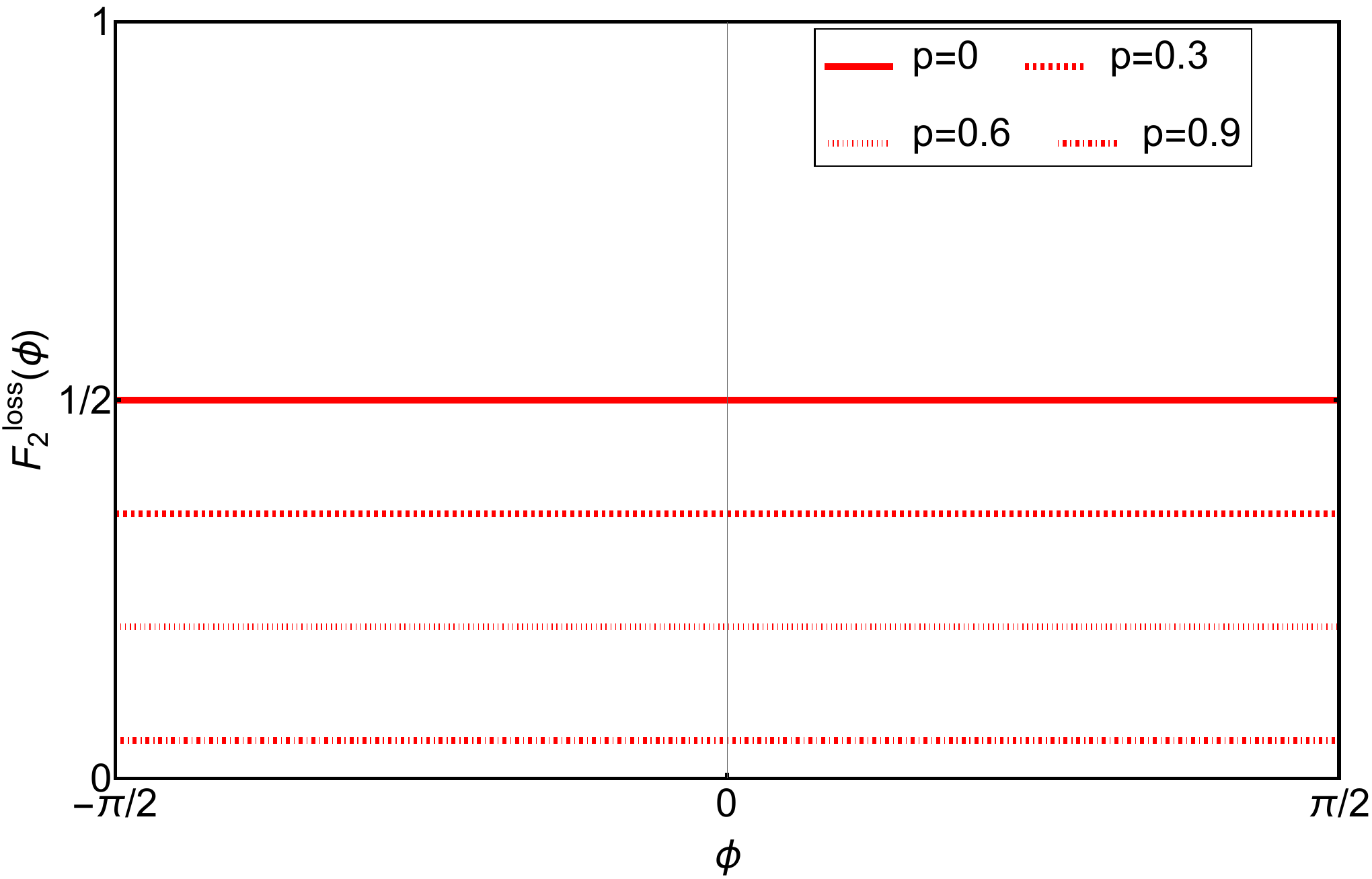}
        \includegraphics[width=3.0in]{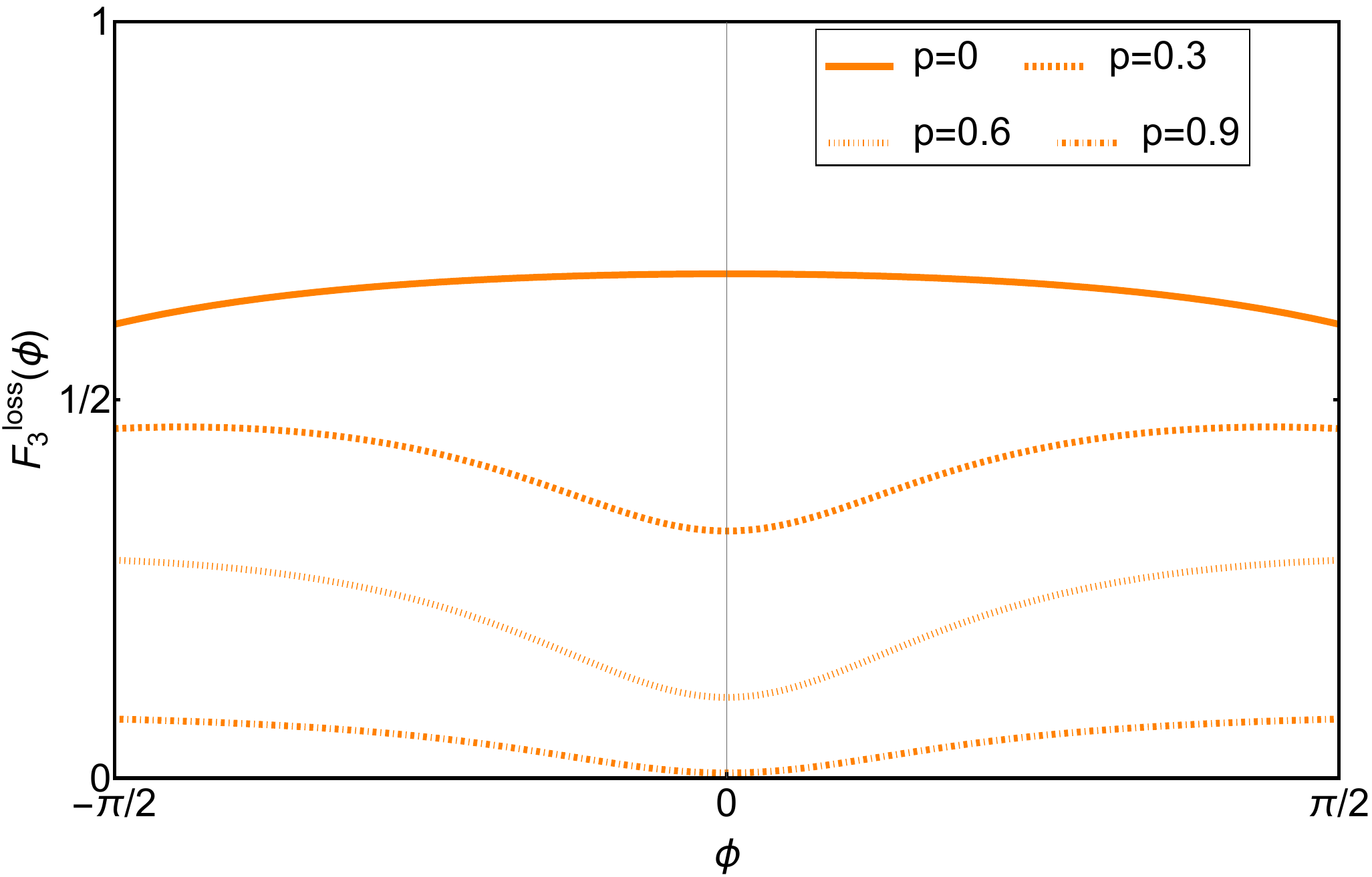}
        \includegraphics[width=3.0in]{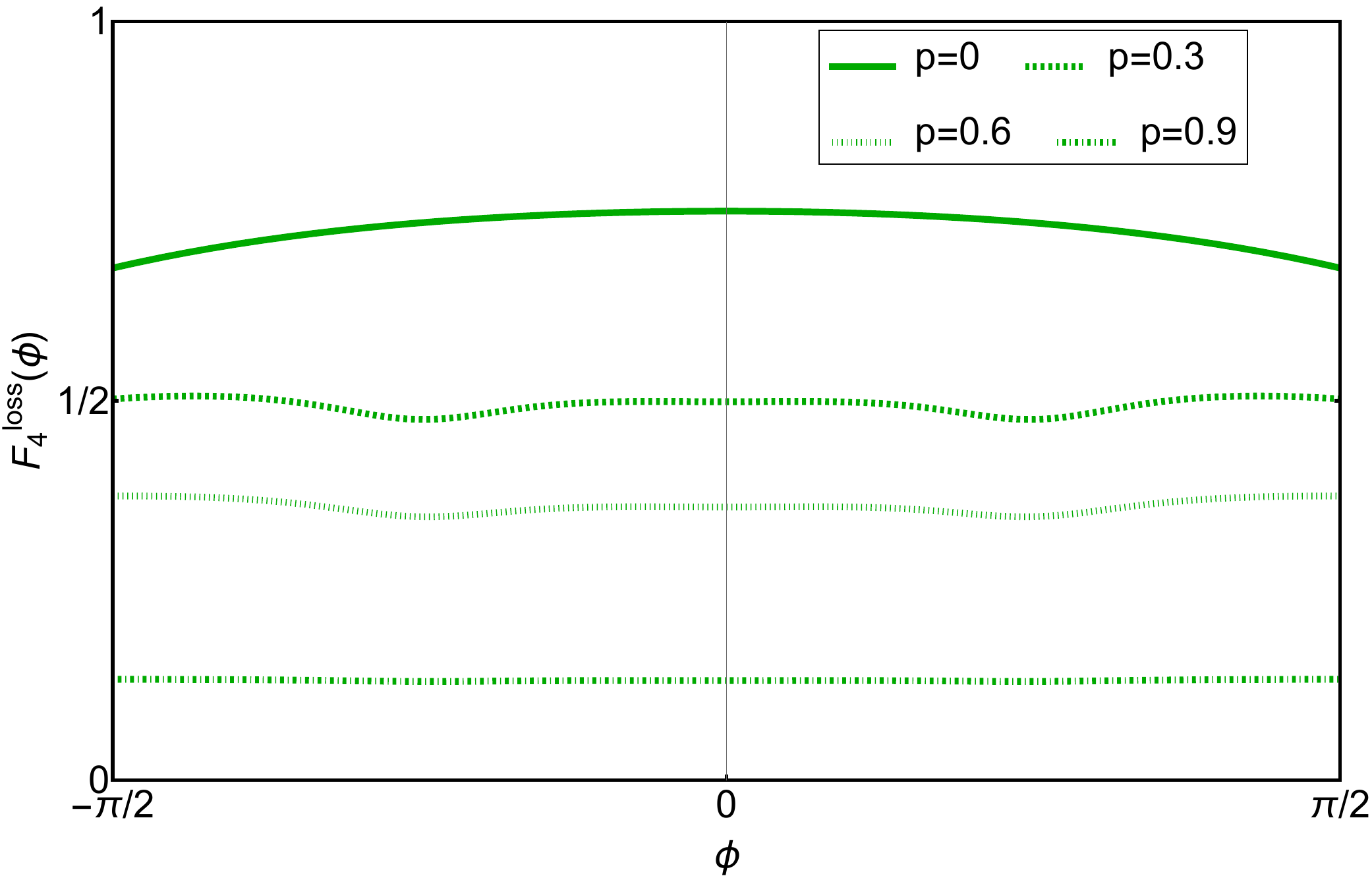}        
        \includegraphics[width=3.0in]{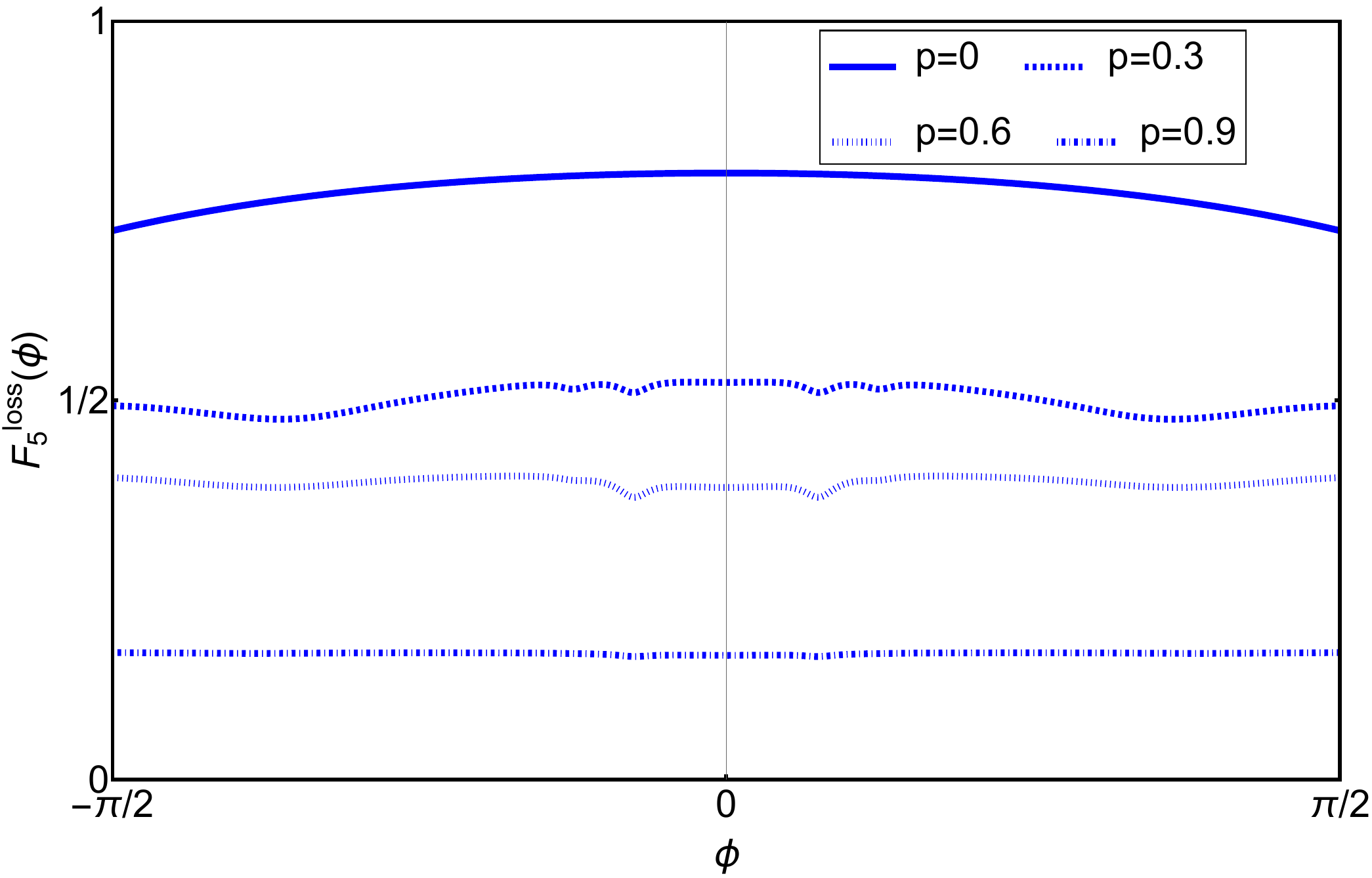}
        \caption{Fisher information in the loss case for $N=2, N=3$, $N=4$, $N=5$  lab-photons. In all cases the more loss causes a lowering of the Fisher information, meaning that the estimation of $\phi$ is going to have a larger error.}
        \label{fig:FisherLoss345}
 \end{minipage}
 \end{figure}

\begin{equation}
    F_N^{\rm{loss}}(\phi)=\sum_{D=1}^N\sum_{j=1}^{m_D}q_D\;r_{D,j}\left(\dfrac{\partial \ln{r_{D,j}}}{\partial \phi}\right)^2.
\end{equation}
Now we can recognize that this is a sum of Fisher information for different values of $D$ weighted with the factor $q_D$
\begin{equation}
    F_N^{\rm{loss}}(\phi)=\sum_{D=1}^N q_D F'_D(\phi),
    \label{eq:FisherD}
\end{equation}
where we called
\begin{equation}
    F'_D(\phi)=\sum_{j=1}^{m_D}r_{D,j}\left(\dfrac{\partial \ln{r_{D,j}}}{\partial \phi}\right)^2.
\end{equation}
Note that these contributions do not correspond to the Fisher information from Eq.~\eqref{eq:FisherN} with a different number of photons.  They are computed by using the probabilities of distributing $D\in[1,N]$ photons in $2N$ detectors, therefore only the first term $F'_N=1-\dfrac{1}{N}$, optimal detection case, coincide with Eq.~\eqref{eq:fisherN_NOLoss}.
The value $q_D$ is the probability of detecting $D=N-k$ photons, and it is given by
\begin{equation}
    q_D=(1-p)^{D-1}(p)^k\;\binom{N-1}{k},
\end{equation}
where $p$ is the probability of losing one photon and the binomial factor arises from the fact that the lost photons are not distinguishable. We point out that $F'_1$ is always zero because in the presence of only one detected photon, $D=1$, we have no information about the correlations. By substituting $D=N-k$ in Eq.~\eqref{eq:FisherD} we recover the expression from Eq.~\eqref{eq:fisherLOSS}.
In Fig.~\ref{fig:FisherLoss345} we present the Fisher information when the photon loss is included. We compare the situations with $N=3$, $N=4$, $N=5$ ground-based photons, and we notice that increasing $p$ results in a lower value of the Fisher information.

\section{Fisher information for lossy case}\noindent
The Fisher information in the no-loss case is given by Eq.~\eqref{eq:FisherN}, where the probabilities $P_\mathbf{d}$ are of the detector signatures $\mathbf{ d}$. In the lossy case, we still have a probability distribution over the detector signatures $\mathbf{ d}$, but the total photon count is no longer always $N$, so generally can be written as 
\begin{equation}
    F(\phi) = \sum_\mathbf{d}^{\sigma_D} P_T(\mathbf{d}) \left( \frac{\partial \ln P_T(\mathbf{d})}{\partial \phi} \right)^2 \,, 
\end{equation} 
where $\sigma_D$ is the outcome set of $D\leq N$ detected photons. By using the chain rule, we obtain Eq.~\eqref{eq:FisherLossD}.


\begin{thebibliography}{39}%
\makeatletter
\providecommand \@ifxundefined [1]{%
 \@ifx{#1\undefined}
}%
\providecommand \@ifnum [1]{%
 \ifnum #1\expandafter \@firstoftwo
 \else \expandafter \@secondoftwo
 \fi
}%
\providecommand \@ifx [1]{%
 \ifx #1\expandafter \@firstoftwo
 \else \expandafter \@secondoftwo
 \fi
}%
\providecommand \natexlab [1]{#1}%
\providecommand \enquote  [1]{``#1''}%
\providecommand \bibnamefont  [1]{#1}%
\providecommand \bibfnamefont [1]{#1}%
\providecommand \citenamefont [1]{#1}%
\providecommand \href@noop [0]{\@secondoftwo}%
\providecommand \href [0]{\begingroup \@sanitize@url \@href}%
\providecommand \@href[1]{\@@startlink{#1}\@@href}%
\providecommand \@@href[1]{\endgroup#1\@@endlink}%
\providecommand \@sanitize@url [0]{\catcode `\\12\catcode `\$12\catcode
  `\&12\catcode `\#12\catcode `\^12\catcode `\_12\catcode `\%12\relax}%
\providecommand \@@startlink[1]{}%
\providecommand \@@endlink[0]{}%
\providecommand \url  [0]{\begingroup\@sanitize@url \@url }%
\providecommand \@url [1]{\endgroup\@href {#1}{\urlprefix }}%
\providecommand \urlprefix  [0]{URL }%
\providecommand \Eprint [0]{\href }%
\providecommand \doibase [0]{http://dx.doi.org/}%
\providecommand \selectlanguage [0]{\@gobble}%
\providecommand \bibinfo  [0]{\@secondoftwo}%
\providecommand \bibfield  [0]{\@secondoftwo}%
\providecommand \translation [1]{[#1]}%
\providecommand \BibitemOpen [0]{}%
\providecommand \bibitemStop [0]{}%
\providecommand \bibitemNoStop [0]{.\EOS\space}%
\providecommand \EOS [0]{\spacefactor3000\relax}%
\providecommand \BibitemShut  [1]{\csname bibitem#1\endcsname}%
\let\auto@bib@innerbib\@empty
\bibitem [{\citenamefont {Pearce}\ \emph {et~al.}(2017)\citenamefont {Pearce},
  \citenamefont {Campbell},\ and\ \citenamefont {Kok}}]{pearce2017optimal}%
  \BibitemOpen
  \bibfield  {author} {\bibinfo {author} {\bibfnamefont {M.~E.}\ \bibnamefont
  {Pearce}}, \bibinfo {author} {\bibfnamefont {E.~T.}\ \bibnamefont
  {Campbell}}, \ and\ \bibinfo {author} {\bibfnamefont {P.}~\bibnamefont
  {Kok}},\ }\href@noop {} {\bibfield  {journal} {\bibinfo  {journal} {Quantum}\
  }\textbf {\bibinfo {volume} {1}},\ \bibinfo {pages} {21} (\bibinfo {year}
  {2017})}\BibitemShut {NoStop}%
\bibitem [{\citenamefont {Howard}\ \emph {et~al.}(2019)\citenamefont {Howard},
  \citenamefont {Gillett}, \citenamefont {Pearce}, \citenamefont {Abrahao},
  \citenamefont {Weinhold}, \citenamefont {Kok},\ and\ \citenamefont
  {White}}]{howard2019}%
  \BibitemOpen
  \bibfield  {author} {\bibinfo {author} {\bibfnamefont {L.~A.}\ \bibnamefont
  {Howard}}, \bibinfo {author} {\bibfnamefont {G.~G.}\ \bibnamefont {Gillett}},
  \bibinfo {author} {\bibfnamefont {M.~E.}\ \bibnamefont {Pearce}}, \bibinfo
  {author} {\bibfnamefont {R.~A.}\ \bibnamefont {Abrahao}}, \bibinfo {author}
  {\bibfnamefont {T.~J.}\ \bibnamefont {Weinhold}}, \bibinfo {author}
  {\bibfnamefont {P.}~\bibnamefont {Kok}}, \ and\ \bibinfo {author}
  {\bibfnamefont {A.~G.}\ \bibnamefont {White}},\ }\href {\doibase
  10.1103/PhysRevLett.123.143604} {\bibfield  {journal} {\bibinfo  {journal}
  {{Phys.\ Rev.\ Lett.}}\ }\textbf {\bibinfo {volume} {123}},\ \bibinfo {pages}
  {143604} (\bibinfo {year} {2019})}\BibitemShut {NoStop}%
\bibitem [{\citenamefont {Kolobov}(2007)}]{kolobov2007quantum}%
  \BibitemOpen
  \bibfield  {author} {\bibinfo {author} {\bibfnamefont {M.}~\bibnamefont
  {Kolobov}},\ }\href@noop {} {\emph {\bibinfo {title} {Quantum Imaging}}}\
  (\bibinfo  {publisher} {Springer New York},\ \bibinfo {year}
  {2007})\BibitemShut {NoStop}%
\bibitem [{\citenamefont {Tsang}\ \emph {et~al.}(2016)\citenamefont {Tsang},
  \citenamefont {Nair},\ and\ \citenamefont {Lu}}]{tsang2016quantum}%
  \BibitemOpen
  \bibfield  {author} {\bibinfo {author} {\bibfnamefont {M.}~\bibnamefont
  {Tsang}}, \bibinfo {author} {\bibfnamefont {R.}~\bibnamefont {Nair}}, \ and\
  \bibinfo {author} {\bibfnamefont {X.-M.}\ \bibnamefont {Lu}},\ }\href@noop {}
  {\bibfield  {journal} {\bibinfo  {journal} {Physical Review X}\ }\textbf
  {\bibinfo {volume} {6}},\ \bibinfo {pages} {031033} (\bibinfo {year}
  {2016})}\BibitemShut {NoStop}%
\bibitem [{\citenamefont {Pa{\'u}r}\ \emph {et~al.}(2016)\citenamefont
  {Pa{\'u}r}, \citenamefont {Stoklasa}, \citenamefont {Hradil}, \citenamefont
  {S{\'a}nchez-Soto},\ and\ \citenamefont {Rehacek}}]{paur2016achieving}%
  \BibitemOpen
  \bibfield  {author} {\bibinfo {author} {\bibfnamefont {M.}~\bibnamefont
  {Pa{\'u}r}}, \bibinfo {author} {\bibfnamefont {B.}~\bibnamefont {Stoklasa}},
  \bibinfo {author} {\bibfnamefont {Z.}~\bibnamefont {Hradil}}, \bibinfo
  {author} {\bibfnamefont {L.~L.}\ \bibnamefont {S{\'a}nchez-Soto}}, \ and\
  \bibinfo {author} {\bibfnamefont {J.}~\bibnamefont {Rehacek}},\ }\href@noop
  {} {\bibfield  {journal} {\bibinfo  {journal} {Optica}\ }\textbf {\bibinfo
  {volume} {3}},\ \bibinfo {pages} {1144} (\bibinfo {year} {2016})}\BibitemShut
  {NoStop}%
\bibitem [{\citenamefont {Tsang}(2019)}]{tsang2019resolving}%
  \BibitemOpen
  \bibfield  {author} {\bibinfo {author} {\bibfnamefont {M.}~\bibnamefont
  {Tsang}},\ }\href@noop {} {\bibfield  {journal} {\bibinfo  {journal}
  {Contemporary Physics}\ }\textbf {\bibinfo {volume} {60}},\ \bibinfo {pages}
  {279} (\bibinfo {year} {2019})}\BibitemShut {NoStop}%
\bibitem [{\citenamefont {Lupo}\ \emph {et~al.}(2020)\citenamefont {Lupo},
  \citenamefont {Huang},\ and\ \citenamefont {Kok}}]{lupo2020}%
  \BibitemOpen
  \bibfield  {author} {\bibinfo {author} {\bibfnamefont {C.}~\bibnamefont
  {Lupo}}, \bibinfo {author} {\bibfnamefont {Z.}~\bibnamefont {Huang}}, \ and\
  \bibinfo {author} {\bibfnamefont {P.}~\bibnamefont {Kok}},\ }\href {\doibase
  10.1103/PhysRevLett.124.080503} {\bibfield  {journal} {\bibinfo  {journal}
  {{Phys.\ Rev.\ Lett.}}\ }\textbf {\bibinfo {volume} {124}},\ \bibinfo {pages}
  {080503} (\bibinfo {year} {2020})}\BibitemShut {NoStop}%
\bibitem [{\citenamefont {Zanforlin}\ \emph {et~al.}(2022)\citenamefont
  {Zanforlin}, \citenamefont {Lupo}, \citenamefont {Connolly}, \citenamefont
  {Kok}, \citenamefont {Buller},\ and\ \citenamefont {Huang}}]{zanforlin2022}%
  \BibitemOpen
  \bibfield  {author} {\bibinfo {author} {\bibfnamefont {U.}~\bibnamefont
  {Zanforlin}}, \bibinfo {author} {\bibfnamefont {C.}~\bibnamefont {Lupo}},
  \bibinfo {author} {\bibfnamefont {P.~W.~R.}\ \bibnamefont {Connolly}},
  \bibinfo {author} {\bibfnamefont {P.}~\bibnamefont {Kok}}, \bibinfo {author}
  {\bibfnamefont {G.~S.}\ \bibnamefont {Buller}}, \ and\ \bibinfo {author}
  {\bibfnamefont {Z.}~\bibnamefont {Huang}},\ }\href {\doibase
  10.1038/s41467-022-32977-8} {\bibfield  {journal} {\bibinfo  {journal}
  {{Nature Comms.}}\ }\textbf {\bibinfo {volume} {13}},\ \bibinfo {pages}
  {5373} (\bibinfo {year} {2022})}\BibitemShut {NoStop}%
\bibitem [{\citenamefont {Brown}\ \emph
  {et~al.}(2022{\natexlab{a}})\citenamefont {Brown}, \citenamefont {Allgaier},
  \citenamefont {Thiel}, \citenamefont {Monnier}, \citenamefont {Raymer},\ and\
  \citenamefont {Smith}}]{Raymer22}%
  \BibitemOpen
  \bibfield  {author} {\bibinfo {author} {\bibfnamefont {M.~R.}\ \bibnamefont
  {Brown}}, \bibinfo {author} {\bibfnamefont {M.}~\bibnamefont {Allgaier}},
  \bibinfo {author} {\bibfnamefont {V.}~\bibnamefont {Thiel}}, \bibinfo
  {author} {\bibfnamefont {J.}~\bibnamefont {Monnier}}, \bibinfo {author}
  {\bibfnamefont {M.~G.}\ \bibnamefont {Raymer}}, \ and\ \bibinfo {author}
  {\bibfnamefont {B.~J.}\ \bibnamefont {Smith}},\ }\href@noop {} {\bibfield
  {journal} {\bibinfo  {journal} {arXiv:2212.07395}\ } (\bibinfo {year}
  {2022}{\natexlab{a}})}\BibitemShut {NoStop}%
\bibitem [{\citenamefont {Dravins}(2016)}]{Dravins16}%
  \BibitemOpen
  \bibfield  {author} {\bibinfo {author} {\bibfnamefont {D.}~\bibnamefont
  {Dravins}},\ }\href {\doibase 10.1117/12.2234130} {\bibfield  {journal}
  {\bibinfo  {journal} {{Proc. SPIE, Optical and Infrared Interferometry and
  Imaging V}}\ }\textbf {\bibinfo {volume} {9907}},\ \bibinfo {pages} {0M}
  (\bibinfo {year} {2016})}\BibitemShut {NoStop}%
\bibitem [{\citenamefont {Akiyama}\ \emph {et~al.}(2019)\citenamefont
  {Akiyama}, \citenamefont {Alberdi}, \citenamefont {Alef}, \citenamefont
  {Asada}, \citenamefont {Azulay}, \citenamefont {Baczko}, \citenamefont
  {Ball}, \citenamefont {Balokovi{\'c}}, \citenamefont {Barrett}, \citenamefont
  {Bintley} \emph {et~al.}}]{akiyama2019first}%
  \BibitemOpen
  \bibfield  {author} {\bibinfo {author} {\bibfnamefont {K.}~\bibnamefont
  {Akiyama}}, \bibinfo {author} {\bibfnamefont {A.}~\bibnamefont {Alberdi}},
  \bibinfo {author} {\bibfnamefont {W.}~\bibnamefont {Alef}}, \bibinfo {author}
  {\bibfnamefont {K.}~\bibnamefont {Asada}}, \bibinfo {author} {\bibfnamefont
  {R.}~\bibnamefont {Azulay}}, \bibinfo {author} {\bibfnamefont {A.-K.}\
  \bibnamefont {Baczko}}, \bibinfo {author} {\bibfnamefont {D.}~\bibnamefont
  {Ball}}, \bibinfo {author} {\bibfnamefont {M.}~\bibnamefont {Balokovi{\'c}}},
  \bibinfo {author} {\bibfnamefont {J.}~\bibnamefont {Barrett}}, \bibinfo
  {author} {\bibfnamefont {D.}~\bibnamefont {Bintley}},  \emph {et~al.},\
  }\href@noop {} {\bibfield  {journal} {\bibinfo  {journal} {The Astrophysical
  Journal Letters}\ }\textbf {\bibinfo {volume} {875}},\ \bibinfo {pages} {L4}
  (\bibinfo {year} {2019})}\BibitemShut {NoStop}%
\bibitem [{\citenamefont {Bojer}\ \emph {et~al.}(2020)\citenamefont {Bojer},
  \citenamefont {Huang}, \citenamefont {Karl}, \citenamefont {Richter},
  \citenamefont {Kok},\ and\ \citenamefont {von Zanthier}}]{boyer2022}%
  \BibitemOpen
  \bibfield  {author} {\bibinfo {author} {\bibfnamefont {M.}~\bibnamefont
  {Bojer}}, \bibinfo {author} {\bibfnamefont {Z.}~\bibnamefont {Huang}},
  \bibinfo {author} {\bibfnamefont {S.}~\bibnamefont {Karl}}, \bibinfo {author}
  {\bibfnamefont {S.}~\bibnamefont {Richter}}, \bibinfo {author} {\bibfnamefont
  {P.}~\bibnamefont {Kok}}, \ and\ \bibinfo {author} {\bibfnamefont
  {J.}~\bibnamefont {von Zanthier}},\ }\href {\doibase
  10.1088/1367-2630/ac5f30} {\bibfield  {journal} {\bibinfo  {journal} {New J.\
  Phys.}\ }\textbf {\bibinfo {volume} {24}},\ \bibinfo {pages} {043026}
  (\bibinfo {year} {2020})}\BibitemShut {NoStop}%
\bibitem [{\citenamefont {Czupryniak}\ \emph {et~al.}(2021)\citenamefont
  {Czupryniak}, \citenamefont {Steinmetz}, \citenamefont {Kwiat},\ and\
  \citenamefont {Jordan}}]{Czupryniak2021}%
  \BibitemOpen
  \bibfield  {author} {\bibinfo {author} {\bibfnamefont {R.}~\bibnamefont
  {Czupryniak}}, \bibinfo {author} {\bibfnamefont {J.}~\bibnamefont
  {Steinmetz}}, \bibinfo {author} {\bibfnamefont {P.~G.}\ \bibnamefont
  {Kwiat}}, \ and\ \bibinfo {author} {\bibfnamefont {A.~N.}\ \bibnamefont
  {Jordan}},\ }\href {\doibase https://doi.org/10.48550/arXiv.2108.01170}
  {\bibfield  {journal} {\bibinfo  {journal} {arXiv:2108.01170}\ } (\bibinfo
  {year} {2021}),\ https://doi.org/10.48550/arXiv.2108.01170}\BibitemShut
  {NoStop}%
\bibitem [{\citenamefont {Brown}\ \emph
  {et~al.}(2022{\natexlab{b}})\citenamefont {Brown}, \citenamefont {Allgaier},
  \citenamefont {Thiel}, \citenamefont {Monnier}, \citenamefont {Raymer}, ,\
  and\ \citenamefont {Smith}}]{Raymer2022}%
  \BibitemOpen
  \bibfield  {author} {\bibinfo {author} {\bibfnamefont {M.~R.}\ \bibnamefont
  {Brown}}, \bibinfo {author} {\bibfnamefont {M.}~\bibnamefont {Allgaier}},
  \bibinfo {author} {\bibfnamefont {V.}~\bibnamefont {Thiel}}, \bibinfo
  {author} {\bibfnamefont {J.}~\bibnamefont {Monnier}}, \bibinfo {author}
  {\bibfnamefont {M.~G.}\ \bibnamefont {Raymer}}, , \ and\ \bibinfo {author}
  {\bibfnamefont {B.~J.}\ \bibnamefont {Smith}},\ }\href {\doibase
  https://doi.org/10.48550/arXiv.2212.07395} {\bibfield  {journal} {\bibinfo
  {journal} {arXiv:2212.07395}\ } (\bibinfo {year} {2022}{\natexlab{b}}),\
  https://doi.org/10.48550/arXiv.2212.07395}\BibitemShut {NoStop}%
\bibitem [{\citenamefont {Townsend}\ \emph {et~al.}(1993)\citenamefont
  {Townsend}, \citenamefont {Rarity},\ and\ \citenamefont
  {Tapster}}]{townsend1993single}%
  \BibitemOpen
  \bibfield  {author} {\bibinfo {author} {\bibfnamefont {P.~D.}\ \bibnamefont
  {Townsend}}, \bibinfo {author} {\bibfnamefont {J.}~\bibnamefont {Rarity}}, \
  and\ \bibinfo {author} {\bibfnamefont {P.}~\bibnamefont {Tapster}},\
  }\href@noop {} {\bibfield  {journal} {\bibinfo  {journal} {Electronics
  Letters}\ }\textbf {\bibinfo {volume} {7}},\ \bibinfo {pages} {634} (\bibinfo
  {year} {1993})}\BibitemShut {NoStop}%
\bibitem [{\citenamefont {Gottesman}\ \emph {et~al.}(2012)\citenamefont
  {Gottesman}, \citenamefont {Jennewein},\ and\ \citenamefont
  {Croke}}]{gottesman2012longer}%
  \BibitemOpen
  \bibfield  {author} {\bibinfo {author} {\bibfnamefont {D.}~\bibnamefont
  {Gottesman}}, \bibinfo {author} {\bibfnamefont {T.}~\bibnamefont
  {Jennewein}}, \ and\ \bibinfo {author} {\bibfnamefont {S.}~\bibnamefont
  {Croke}},\ }\href@noop {} {\bibfield  {journal} {\bibinfo  {journal}
  {Physical review letters}\ }\textbf {\bibinfo {volume} {109}},\ \bibinfo
  {pages} {070503} (\bibinfo {year} {2012})}\BibitemShut {NoStop}%
\bibitem [{\citenamefont {Khabiboulline}\ \emph {et~al.}(2019)\citenamefont
  {Khabiboulline}, \citenamefont {Borregaard}, \citenamefont {De~Greve},\ and\
  \citenamefont {Lukin}}]{khabiboulline2019optical}%
  \BibitemOpen
  \bibfield  {author} {\bibinfo {author} {\bibfnamefont {E.~T.}\ \bibnamefont
  {Khabiboulline}}, \bibinfo {author} {\bibfnamefont {J.}~\bibnamefont
  {Borregaard}}, \bibinfo {author} {\bibfnamefont {K.}~\bibnamefont
  {De~Greve}}, \ and\ \bibinfo {author} {\bibfnamefont {M.~D.}\ \bibnamefont
  {Lukin}},\ }\href@noop {} {\bibfield  {journal} {\bibinfo  {journal}
  {Physical review letters}\ }\textbf {\bibinfo {volume} {123}},\ \bibinfo
  {pages} {070504} (\bibinfo {year} {2019})}\BibitemShut {NoStop}%
\bibitem [{\citenamefont {Huang}\ \emph {et~al.}(2022)\citenamefont {Huang},
  \citenamefont {Brennen},\ and\ \citenamefont
  {Ouyang}}]{PhysRevLett.129.210502}%
  \BibitemOpen
  \bibfield  {author} {\bibinfo {author} {\bibfnamefont {Z.}~\bibnamefont
  {Huang}}, \bibinfo {author} {\bibfnamefont {G.~K.}\ \bibnamefont {Brennen}},
  \ and\ \bibinfo {author} {\bibfnamefont {Y.}~\bibnamefont {Ouyang}},\ }\href
  {\doibase 10.1103/PhysRevLett.129.210502} {\bibfield  {journal} {\bibinfo
  {journal} {Phys. Rev. Lett.}\ }\textbf {\bibinfo {volume} {129}},\ \bibinfo
  {pages} {210502} (\bibinfo {year} {2022})}\BibitemShut {NoStop}%
\bibitem [{\citenamefont {Kaneda}\ and\ \citenamefont
  {Kwiat}(2019)}]{kaneda2019high}%
  \BibitemOpen
  \bibfield  {author} {\bibinfo {author} {\bibfnamefont {F.}~\bibnamefont
  {Kaneda}}\ and\ \bibinfo {author} {\bibfnamefont {P.~G.}\ \bibnamefont
  {Kwiat}},\ }\href@noop {} {\bibfield  {journal} {\bibinfo  {journal} {Science
  advances}\ }\textbf {\bibinfo {volume} {5}},\ \bibinfo {pages} {eaaw8586}
  (\bibinfo {year} {2019})}\BibitemShut {NoStop}%
\bibitem [{\citenamefont {Kennard}\ \emph {et~al.}(2013)\citenamefont
  {Kennard}, \citenamefont {Hadden}, \citenamefont {Marseglia}, \citenamefont
  {Aharonovich}, \citenamefont {Castelletto}, \citenamefont {Patton},
  \citenamefont {Politi}, \citenamefont {Matthews}, \citenamefont {Sinclair},
  \citenamefont {Gibson} \emph {et~al.}}]{kennard2013chip}%
  \BibitemOpen
  \bibfield  {author} {\bibinfo {author} {\bibfnamefont {J.}~\bibnamefont
  {Kennard}}, \bibinfo {author} {\bibfnamefont {J.}~\bibnamefont {Hadden}},
  \bibinfo {author} {\bibfnamefont {L.}~\bibnamefont {Marseglia}}, \bibinfo
  {author} {\bibfnamefont {I.}~\bibnamefont {Aharonovich}}, \bibinfo {author}
  {\bibfnamefont {S.}~\bibnamefont {Castelletto}}, \bibinfo {author}
  {\bibfnamefont {B.}~\bibnamefont {Patton}}, \bibinfo {author} {\bibfnamefont
  {A.}~\bibnamefont {Politi}}, \bibinfo {author} {\bibfnamefont
  {J.}~\bibnamefont {Matthews}}, \bibinfo {author} {\bibfnamefont
  {A.}~\bibnamefont {Sinclair}}, \bibinfo {author} {\bibfnamefont
  {B.}~\bibnamefont {Gibson}},  \emph {et~al.},\ }\href@noop {} {\bibfield
  {journal} {\bibinfo  {journal} {Physical review letters}\ }\textbf {\bibinfo
  {volume} {111}},\ \bibinfo {pages} {213603} (\bibinfo {year}
  {2013})}\BibitemShut {NoStop}%
\bibitem [{\citenamefont {Fulconis}\ \emph {et~al.}(2005)\citenamefont
  {Fulconis}, \citenamefont {Alibart}, \citenamefont {Wadsworth}, \citenamefont
  {Russell},\ and\ \citenamefont {Rarity}}]{fulconis2005high}%
  \BibitemOpen
  \bibfield  {author} {\bibinfo {author} {\bibfnamefont {J.}~\bibnamefont
  {Fulconis}}, \bibinfo {author} {\bibfnamefont {O.}~\bibnamefont {Alibart}},
  \bibinfo {author} {\bibfnamefont {W.}~\bibnamefont {Wadsworth}}, \bibinfo
  {author} {\bibfnamefont {P.~S.~J.}\ \bibnamefont {Russell}}, \ and\ \bibinfo
  {author} {\bibfnamefont {J.}~\bibnamefont {Rarity}},\ }\href@noop {}
  {\bibfield  {journal} {\bibinfo  {journal} {Optics Express}\ }\textbf
  {\bibinfo {volume} {13}},\ \bibinfo {pages} {7572} (\bibinfo {year}
  {2005})}\BibitemShut {NoStop}%
\bibitem [{\citenamefont {Silverstone}\ \emph {et~al.}(2014)\citenamefont
  {Silverstone}, \citenamefont {Bonneau}, \citenamefont {Ohira}, \citenamefont
  {Suzuki}, \citenamefont {Yoshida}, \citenamefont {Iizuka}, \citenamefont
  {Ezaki}, \citenamefont {Natarajan}, \citenamefont {Tanner}, \citenamefont
  {Hadfield} \emph {et~al.}}]{silverstone2014chip}%
  \BibitemOpen
  \bibfield  {author} {\bibinfo {author} {\bibfnamefont {J.~W.}\ \bibnamefont
  {Silverstone}}, \bibinfo {author} {\bibfnamefont {D.}~\bibnamefont
  {Bonneau}}, \bibinfo {author} {\bibfnamefont {K.}~\bibnamefont {Ohira}},
  \bibinfo {author} {\bibfnamefont {N.}~\bibnamefont {Suzuki}}, \bibinfo
  {author} {\bibfnamefont {H.}~\bibnamefont {Yoshida}}, \bibinfo {author}
  {\bibfnamefont {N.}~\bibnamefont {Iizuka}}, \bibinfo {author} {\bibfnamefont
  {M.}~\bibnamefont {Ezaki}}, \bibinfo {author} {\bibfnamefont {C.~M.}\
  \bibnamefont {Natarajan}}, \bibinfo {author} {\bibfnamefont {M.~G.}\
  \bibnamefont {Tanner}}, \bibinfo {author} {\bibfnamefont {R.~H.}\
  \bibnamefont {Hadfield}},  \emph {et~al.},\ }\href@noop {} {\bibfield
  {journal} {\bibinfo  {journal} {Nature Photonics}\ }\textbf {\bibinfo
  {volume} {8}},\ \bibinfo {pages} {104} (\bibinfo {year} {2014})}\BibitemShut
  {NoStop}%
\bibitem [{\citenamefont {Wang}\ \emph {et~al.}(2020)\citenamefont {Wang},
  \citenamefont {Sciarrino}, \citenamefont {Laing},\ and\ \citenamefont
  {Thompson}}]{wang2020integrated}%
  \BibitemOpen
  \bibfield  {author} {\bibinfo {author} {\bibfnamefont {J.}~\bibnamefont
  {Wang}}, \bibinfo {author} {\bibfnamefont {F.}~\bibnamefont {Sciarrino}},
  \bibinfo {author} {\bibfnamefont {A.}~\bibnamefont {Laing}}, \ and\ \bibinfo
  {author} {\bibfnamefont {M.~G.}\ \bibnamefont {Thompson}},\ }\href@noop {}
  {\bibfield  {journal} {\bibinfo  {journal} {Nature Photonics}\ }\textbf
  {\bibinfo {volume} {14}},\ \bibinfo {pages} {273} (\bibinfo {year}
  {2020})}\BibitemShut {NoStop}%
\bibitem [{\citenamefont {Carolan}\ \emph {et~al.}(2015)\citenamefont
  {Carolan}, \citenamefont {Harrold}, \citenamefont {Sparrow}, \citenamefont
  {Mart{\'\i}n-L{\'o}pez}, \citenamefont {Russell}, \citenamefont
  {Silverstone}, \citenamefont {Shadbolt}, \citenamefont {Matsuda},
  \citenamefont {Oguma}, \citenamefont {Itoh} \emph
  {et~al.}}]{carolan2015universal}%
  \BibitemOpen
  \bibfield  {author} {\bibinfo {author} {\bibfnamefont {J.}~\bibnamefont
  {Carolan}}, \bibinfo {author} {\bibfnamefont {C.}~\bibnamefont {Harrold}},
  \bibinfo {author} {\bibfnamefont {C.}~\bibnamefont {Sparrow}}, \bibinfo
  {author} {\bibfnamefont {E.}~\bibnamefont {Mart{\'\i}n-L{\'o}pez}}, \bibinfo
  {author} {\bibfnamefont {N.~J.}\ \bibnamefont {Russell}}, \bibinfo {author}
  {\bibfnamefont {J.~W.}\ \bibnamefont {Silverstone}}, \bibinfo {author}
  {\bibfnamefont {P.~J.}\ \bibnamefont {Shadbolt}}, \bibinfo {author}
  {\bibfnamefont {N.}~\bibnamefont {Matsuda}}, \bibinfo {author} {\bibfnamefont
  {M.}~\bibnamefont {Oguma}}, \bibinfo {author} {\bibfnamefont
  {M.}~\bibnamefont {Itoh}},  \emph {et~al.},\ }\href@noop {} {\bibfield
  {journal} {\bibinfo  {journal} {Science}\ }\textbf {\bibinfo {volume}
  {349}},\ \bibinfo {pages} {711} (\bibinfo {year} {2015})}\BibitemShut
  {NoStop}%
\bibitem [{\citenamefont {McMillan}\ \emph {et~al.}(2013)\citenamefont
  {McMillan}, \citenamefont {Bell}, \citenamefont {Clark}, \citenamefont
  {Labont{\'e}}, \citenamefont {Kannan}, \citenamefont {McCutcheon},
  \citenamefont {Wu}, \citenamefont {Martin}, \citenamefont {Alibart},
  \citenamefont {Tanzilli} \emph {et~al.}}]{mcmillan2013optical}%
  \BibitemOpen
  \bibfield  {author} {\bibinfo {author} {\bibfnamefont {A.}~\bibnamefont
  {McMillan}}, \bibinfo {author} {\bibfnamefont {B.}~\bibnamefont {Bell}},
  \bibinfo {author} {\bibfnamefont {A.}~\bibnamefont {Clark}}, \bibinfo
  {author} {\bibfnamefont {L.}~\bibnamefont {Labont{\'e}}}, \bibinfo {author}
  {\bibfnamefont {S.}~\bibnamefont {Kannan}}, \bibinfo {author} {\bibfnamefont
  {W.}~\bibnamefont {McCutcheon}}, \bibinfo {author} {\bibfnamefont
  {T.}~\bibnamefont {Wu}}, \bibinfo {author} {\bibfnamefont {A.}~\bibnamefont
  {Martin}}, \bibinfo {author} {\bibfnamefont {O.}~\bibnamefont {Alibart}},
  \bibinfo {author} {\bibfnamefont {S.}~\bibnamefont {Tanzilli}},  \emph
  {et~al.},\ }in\ \href@noop {} {\emph {\bibinfo {booktitle} {Laser Science}}}\
  (\bibinfo {organization} {Optica Publishing Group},\ \bibinfo {year} {2013})\
  pp.\ \bibinfo {pages} {LTu4G--3}\BibitemShut {NoStop}%
\bibitem [{\citenamefont {Paesani}\ \emph {et~al.}(2020)\citenamefont
  {Paesani}, \citenamefont {Borghi}, \citenamefont {Signorini}, \citenamefont
  {Ma{\"\i}nos}, \citenamefont {Pavesi},\ and\ \citenamefont
  {Laing}}]{paesani2020near}%
  \BibitemOpen
  \bibfield  {author} {\bibinfo {author} {\bibfnamefont {S.}~\bibnamefont
  {Paesani}}, \bibinfo {author} {\bibfnamefont {M.}~\bibnamefont {Borghi}},
  \bibinfo {author} {\bibfnamefont {S.}~\bibnamefont {Signorini}}, \bibinfo
  {author} {\bibfnamefont {A.}~\bibnamefont {Ma{\"\i}nos}}, \bibinfo {author}
  {\bibfnamefont {L.}~\bibnamefont {Pavesi}}, \ and\ \bibinfo {author}
  {\bibfnamefont {A.}~\bibnamefont {Laing}},\ }\href@noop {} {\bibfield
  {journal} {\bibinfo  {journal} {Nature communications}\ }\textbf {\bibinfo
  {volume} {11}},\ \bibinfo {pages} {1} (\bibinfo {year} {2020})}\BibitemShut
  {NoStop}%
\bibitem [{\citenamefont {Hadfield}(2009)}]{hadfield2009single}%
  \BibitemOpen
  \bibfield  {author} {\bibinfo {author} {\bibfnamefont {R.~H.}\ \bibnamefont
  {Hadfield}},\ }\href@noop {} {\bibfield  {journal} {\bibinfo  {journal}
  {Nature photonics}\ }\textbf {\bibinfo {volume} {3}},\ \bibinfo {pages} {696}
  (\bibinfo {year} {2009})}\BibitemShut {NoStop}%
\bibitem [{\citenamefont {Esmaeil~Zadeh}\ \emph {et~al.}(2021)\citenamefont
  {Esmaeil~Zadeh}, \citenamefont {Chang}, \citenamefont {Los}, \citenamefont
  {Gyger}, \citenamefont {Elshaari}, \citenamefont {Steinhauer}, \citenamefont
  {Dorenbos},\ and\ \citenamefont {Zwiller}}]{esmaeil2021superconducting}%
  \BibitemOpen
  \bibfield  {author} {\bibinfo {author} {\bibfnamefont {I.}~\bibnamefont
  {Esmaeil~Zadeh}}, \bibinfo {author} {\bibfnamefont {J.}~\bibnamefont
  {Chang}}, \bibinfo {author} {\bibfnamefont {J.~W.}\ \bibnamefont {Los}},
  \bibinfo {author} {\bibfnamefont {S.}~\bibnamefont {Gyger}}, \bibinfo
  {author} {\bibfnamefont {A.~W.}\ \bibnamefont {Elshaari}}, \bibinfo {author}
  {\bibfnamefont {S.}~\bibnamefont {Steinhauer}}, \bibinfo {author}
  {\bibfnamefont {S.~N.}\ \bibnamefont {Dorenbos}}, \ and\ \bibinfo {author}
  {\bibfnamefont {V.}~\bibnamefont {Zwiller}},\ }\href@noop {} {\bibfield
  {journal} {\bibinfo  {journal} {Applied Physics Letters}\ }\textbf {\bibinfo
  {volume} {118}},\ \bibinfo {pages} {190502} (\bibinfo {year}
  {2021})}\BibitemShut {NoStop}%
\bibitem [{\citenamefont {Natarajan}\ \emph {et~al.}(2012)\citenamefont
  {Natarajan}, \citenamefont {Tanner},\ and\ \citenamefont
  {Hadfield}}]{natarajan2012superconducting}%
  \BibitemOpen
  \bibfield  {author} {\bibinfo {author} {\bibfnamefont {C.~M.}\ \bibnamefont
  {Natarajan}}, \bibinfo {author} {\bibfnamefont {M.~G.}\ \bibnamefont
  {Tanner}}, \ and\ \bibinfo {author} {\bibfnamefont {R.~H.}\ \bibnamefont
  {Hadfield}},\ }\href@noop {} {\bibfield  {journal} {\bibinfo  {journal}
  {Superconductor science and technology}\ }\textbf {\bibinfo {volume} {25}},\
  \bibinfo {pages} {063001} (\bibinfo {year} {2012})}\BibitemShut {NoStop}%
\bibitem [{\citenamefont {Hadfield}\ \emph {et~al.}(2006)\citenamefont
  {Hadfield}, \citenamefont {Habif}, \citenamefont {Schlafer}, \citenamefont
  {Schwall},\ and\ \citenamefont {Nam}}]{hadfield2006quantum}%
  \BibitemOpen
  \bibfield  {author} {\bibinfo {author} {\bibfnamefont {R.~H.}\ \bibnamefont
  {Hadfield}}, \bibinfo {author} {\bibfnamefont {J.~L.}\ \bibnamefont {Habif}},
  \bibinfo {author} {\bibfnamefont {J.}~\bibnamefont {Schlafer}}, \bibinfo
  {author} {\bibfnamefont {R.~E.}\ \bibnamefont {Schwall}}, \ and\ \bibinfo
  {author} {\bibfnamefont {S.~W.}\ \bibnamefont {Nam}},\ }\href@noop {}
  {\bibfield  {journal} {\bibinfo  {journal} {Applied physics letters}\
  }\textbf {\bibinfo {volume} {89}},\ \bibinfo {pages} {241129} (\bibinfo
  {year} {2006})}\BibitemShut {NoStop}%
\bibitem [{\citenamefont {Miller}\ \emph {et~al.}(2003)\citenamefont {Miller},
  \citenamefont {Nam}, \citenamefont {Martinis}, \citenamefont {Sergienko}
  \emph {et~al.}}]{miller2003tungsten}%
  \BibitemOpen
  \bibfield  {author} {\bibinfo {author} {\bibfnamefont {A.~J.}\ \bibnamefont
  {Miller}}, \bibinfo {author} {\bibfnamefont {S.~W.}\ \bibnamefont {Nam}},
  \bibinfo {author} {\bibfnamefont {J.~M.}\ \bibnamefont {Martinis}}, \bibinfo
  {author} {\bibfnamefont {A.~V.}\ \bibnamefont {Sergienko}},  \emph {et~al.},\
  }\href@noop {} {\  (\bibinfo {year} {2003})}\BibitemShut {NoStop}%
\bibitem [{\citenamefont {Paris}(2009)}]{paris2009quantum}%
  \BibitemOpen
  \bibfield  {author} {\bibinfo {author} {\bibfnamefont {M.~G.}\ \bibnamefont
  {Paris}},\ }\href@noop {} {\bibfield  {journal} {\bibinfo  {journal}
  {International Journal of Quantum Information}\ }\textbf {\bibinfo {volume}
  {7}},\ \bibinfo {pages} {125} (\bibinfo {year} {2009})}\BibitemShut {NoStop}%
\bibitem [{\citenamefont {Giovannetti}\ \emph {et~al.}(2006)\citenamefont
  {Giovannetti}, \citenamefont {Lloyd},\ and\ \citenamefont
  {Maccone}}]{giovannetti2006quantum}%
  \BibitemOpen
  \bibfield  {author} {\bibinfo {author} {\bibfnamefont {V.}~\bibnamefont
  {Giovannetti}}, \bibinfo {author} {\bibfnamefont {S.}~\bibnamefont {Lloyd}},
  \ and\ \bibinfo {author} {\bibfnamefont {L.}~\bibnamefont {Maccone}},\
  }\href@noop {} {\bibfield  {journal} {\bibinfo  {journal} {Physical review
  letters}\ }\textbf {\bibinfo {volume} {96}},\ \bibinfo {pages} {010401}
  (\bibinfo {year} {2006})}\BibitemShut {NoStop}%
\bibitem [{\citenamefont {Giovannetti}\ \emph {et~al.}(2011)\citenamefont
  {Giovannetti}, \citenamefont {Lloyd},\ and\ \citenamefont
  {Maccone}}]{giovannetti2011advances}%
  \BibitemOpen
  \bibfield  {author} {\bibinfo {author} {\bibfnamefont {V.}~\bibnamefont
  {Giovannetti}}, \bibinfo {author} {\bibfnamefont {S.}~\bibnamefont {Lloyd}},
  \ and\ \bibinfo {author} {\bibfnamefont {L.}~\bibnamefont {Maccone}},\
  }\href@noop {} {\bibfield  {journal} {\bibinfo  {journal} {Nature photonics}\
  }\textbf {\bibinfo {volume} {5}},\ \bibinfo {pages} {222} (\bibinfo {year}
  {2011})}\BibitemShut {NoStop}%
\bibitem [{\citenamefont {Demkowicz-Dobrzanski}\ \emph
  {et~al.}(2009)\citenamefont {Demkowicz-Dobrzanski}, \citenamefont {Dorner},
  \citenamefont {Smith}, \citenamefont {Lundeen}, \citenamefont {Wasilewski},
  \citenamefont {Banaszek},\ and\ \citenamefont
  {Walmsley}}]{demkowicz2009quantum}%
  \BibitemOpen
  \bibfield  {author} {\bibinfo {author} {\bibfnamefont {R.}~\bibnamefont
  {Demkowicz-Dobrzanski}}, \bibinfo {author} {\bibfnamefont {U.}~\bibnamefont
  {Dorner}}, \bibinfo {author} {\bibfnamefont {B.}~\bibnamefont {Smith}},
  \bibinfo {author} {\bibfnamefont {J.}~\bibnamefont {Lundeen}}, \bibinfo
  {author} {\bibfnamefont {W.}~\bibnamefont {Wasilewski}}, \bibinfo {author}
  {\bibfnamefont {K.}~\bibnamefont {Banaszek}}, \ and\ \bibinfo {author}
  {\bibfnamefont {I.}~\bibnamefont {Walmsley}},\ }\href@noop {} {\bibfield
  {journal} {\bibinfo  {journal} {Physical Review A}\ }\textbf {\bibinfo
  {volume} {80}},\ \bibinfo {pages} {013825} (\bibinfo {year}
  {2009})}\BibitemShut {NoStop}%
\bibitem [{\citenamefont {Knott}\ \emph {et~al.}(2016)\citenamefont {Knott},
  \citenamefont {Proctor}, \citenamefont {Hayes}, \citenamefont {Ralph},
  \citenamefont {Kok},\ and\ \citenamefont {Dunningham}}]{knott2016local}%
  \BibitemOpen
  \bibfield  {author} {\bibinfo {author} {\bibfnamefont {P.~A.}\ \bibnamefont
  {Knott}}, \bibinfo {author} {\bibfnamefont {T.~J.}\ \bibnamefont {Proctor}},
  \bibinfo {author} {\bibfnamefont {A.~J.}\ \bibnamefont {Hayes}}, \bibinfo
  {author} {\bibfnamefont {J.~F.}\ \bibnamefont {Ralph}}, \bibinfo {author}
  {\bibfnamefont {P.}~\bibnamefont {Kok}}, \ and\ \bibinfo {author}
  {\bibfnamefont {J.~A.}\ \bibnamefont {Dunningham}},\ }\href@noop {}
  {\bibfield  {journal} {\bibinfo  {journal} {Physical Review A}\ }\textbf
  {\bibinfo {volume} {94}},\ \bibinfo {pages} {062312} (\bibinfo {year}
  {2016})}\BibitemShut {NoStop}%
\bibitem [{\citenamefont {Helstrom}(1976)}]{helstrom1976quantum}%
  \BibitemOpen
  \bibfield  {author} {\bibinfo {author} {\bibfnamefont {C.}~\bibnamefont
  {Helstrom}},\ }\href@noop {} {\emph {\bibinfo {title} {Quantum Detection and
  Estimation Theory}}},\ Vol.~\bibinfo {volume} {84}\ (\bibinfo  {publisher}
  {New York: Academic},\ \bibinfo {year} {1976})\BibitemShut {NoStop}%
\bibitem [{\citenamefont {Reck}\ \emph {et~al.}(1994)\citenamefont {Reck},
  \citenamefont {Zeilinger}, \citenamefont {Bernstein},\ and\ \citenamefont
  {Bertani}}]{reck1994experimental}%
  \BibitemOpen
  \bibfield  {author} {\bibinfo {author} {\bibfnamefont {M.}~\bibnamefont
  {Reck}}, \bibinfo {author} {\bibfnamefont {A.}~\bibnamefont {Zeilinger}},
  \bibinfo {author} {\bibfnamefont {H.~J.}\ \bibnamefont {Bernstein}}, \ and\
  \bibinfo {author} {\bibfnamefont {P.}~\bibnamefont {Bertani}},\ }\href@noop
  {} {\bibfield  {journal} {\bibinfo  {journal} {Physical review letters}\
  }\textbf {\bibinfo {volume} {73}},\ \bibinfo {pages} {58} (\bibinfo {year}
  {1994})}\BibitemShut {NoStop}%
\bibitem [{\citenamefont {Mandel}\ and\ \citenamefont
  {Wolf}(1995)}]{mandel1995optical}%
  \BibitemOpen
  \bibfield  {author} {\bibinfo {author} {\bibfnamefont {L.}~\bibnamefont
  {Mandel}}\ and\ \bibinfo {author} {\bibfnamefont {E.}~\bibnamefont {Wolf}},\
  }\href@noop {} {\emph {\bibinfo {title} {Optical coherence and quantum
  optics}}}\ (\bibinfo  {publisher} {Cambridge university press},\ \bibinfo
  {year} {1995})\BibitemShut {NoStop}%
\end{thebibliography}
\end{document}